\documentclass[twocolumn]{aastex702}
\usepackage{graphicx}	
\usepackage{amsmath}	
\usepackage{amssymb}
\usepackage{color}
\usepackage{array}
\usepackage{bookmark}
\usepackage{subfigure}
\usepackage{booktabs}
\usepackage{threeparttable}

\def\beq{\begin{eqnarray}}
\def\eeq{\end{eqnarray}}

\begin{document}

\title{Migration Traps as Variability Attractors. I. Optical/UV Signatures of Embedded Stellar-Mass Black Holes in Active Galactic Nucleus Disks}

\correspondingauthor{Tong Liu}
\email{tongliu@xmu.edu.cn}

\author[0009-0008-9726-9431]{Jing-Tong Xing}
\affiliation{Department of Astronomy, Xiamen University, Xiamen, Fujian 361005, China}
\email{kriseme@163.com}

\author[0000-0001-8678-6291]{Tong Liu}
\affiliation{Department of Astronomy, Xiamen University, Xiamen, Fujian 361005, China}
\affiliation{SHAO-XMU Joint Center for Astrophysics, Xiamen University, Xiamen, Fujian 361005, China}
\email{tongliu@xmu.edu.cn}

\author[0000-0002-0771-2153]{Mouyuan Sun}
\affiliation{Department of Astronomy, Xiamen University, Xiamen, Fujian 361005, China}
\affiliation{SHAO-XMU Joint Center for Astrophysics, Xiamen University, Xiamen, Fujian 361005, China}
\email{msun88@xmu.edu.cn}

\author[0000-0002-7329-9344]{Ya-Ping Li}
\affiliation{Shanghai Astronomical Observatory, Chinese Academy of Sciences, Shanghai 200030, China}
\affiliation{SHAO-XMU Joint Center for Astrophysics, Xiamen University, Xiamen, Fujian 361005, China}
\email{liyp@shao.ac.cn}

\author[0009-0005-2801-6594]{Shuying Zhou}
\affiliation{Department of Astronomy, Xiamen University, Xiamen, Fujian 361005, China}
\email{zhoushuying@stu.xmu.edu.cn}

\author[0000-0002-4223-2198]{Zhen-Yi Cai}
\affiliation{Department of Astronomy, University of Science and Technology of China, Hefei, Anhui 230026, China}
\affiliation{School of Astronomy and Space Science, University of Science and Technology of China, Hefei, Anhui 230026, China}
\email{zcai@ustc.edu.cn}

\author[0000-0003-1474-293X]{Da-Bin Lin}
\affiliation{Laboratory for Relativistic Astrophysics, Department of Physics, Guangxi University, Nanning 530004, China}
\email{lindabin@gxu.edu.cn}

\author[0000-0001-9449-9268]{Jian-Min Wang}
\affiliation{Key Laboratory for Particle Astrophysics, Institute of High Energy Physics, Chinese Academy of Sciences, Beijing 100049, China}
\affiliation{School of Astronomy and Space Sciences, University of Chinese Academy of Sciences, Beijing 100049, China}
\affiliation{National Astronomical Observatory of China, Beijing 100020, China}
\email{wangjm@ihep.ac.cn}

\begin{abstract}
We investigate whether embedded stellar-mass black holes (sBHs) in active galactic nucleus (AGN) disks can leave observable optical/UV variability signatures through migration-trap-driven magnetic heating. This mechanism operates when sBHs migrating toward torque-balance radii pile up near migration traps, triggering localized, stochastic magnetic reconnection that heats the disk atmosphere. It is potentially important because it provides a physical source of non-coronal disk heating and directly links optical/UV continuum variability to otherwise hidden compact-object populations. By coupling a one-dimensional sBH population synthesis model with a corona-heated accretion-disk reprocessing variability framework, we show that migration traps concentrate sBHs at preferred radii and generate localized, stochastic reconnection heating. The resulting heating is self-regulated: sBH pile-ups enhance the reconnection rate, while gap opening reduces the local gas density and partially suppresses the reconnection power. This heating produces excess short-timescale optical/UV variability, flattened short-term structure functions, and deviations from the standard $\tau\propto\lambda^{4/3}$ lag-wavelength relation, which describes the time delay between variability at different wavelengths for a standard thin accretion disk. These signatures are strongest at low-to-moderate Eddington ratios, and related observations could provide indirect evidence for embedded compact-object populations in AGN disks.
\end{abstract}


\section{Introduction} \label{sec:intro}

Active galactic nuclei (AGNs) display stochastic variability over a broad range of wavelengths and timescales, from rapid X-ray fluctuations produced in the innermost regions to optical/UV variations associated with the accretion disk \cite[e.g.,][]{2003MNRAS.345.1271V,2013MNRAS.430L..49M,2021Sci...373..789B,2024MNRAS.529.2877M}. In the standard picture, the optical/UV continuum is generated by thermal emission from a geometrically thin, optically thick accretion disk \cite[][]{1973A&A....24..337S}, while a hot compact corona produces high-energy radiation close to the supermassive black hole (SMBH). Because different wavelengths predominantly arise from different disk radii, multi-band continuum variability provides a powerful probe of the disk thermal structure and energy transport. In particular, continuum reverberation mapping uses inter-band time delays to infer the characteristic size and radial temperature profile of the emitting disk \cite[e.g.,][]{1982ApJ...255..419B,1991ApJ...371..541K,1991A&A...249..344C,2021iSci...24j2557C}.

In the classical lamp-post reprocessing scenario, variable X-ray emission from the corona irradiates the disk and drives delayed optical/UV responses at larger radii. This framework naturally predicts that longer-wavelength emission should lag behind shorter-wavelength or X-ray emission by approximately the light-travel time across the disk \cite[e.g.,][]{1991A&A...249..344C,2021iSci...24j2557C}. Although such positive lags are broadly consistent with many reverberation measurements, intensive monitoring campaigns have revealed several persistent discrepancies. Observed optical/UV lags are often larger than those expected from standard thin-disk theory, and the inferred wavelength dependence can deviate from the canonical $\tau \propto \lambda^{4/3}$ scaling \cite[e.g.,][]{2015ApJ...806..129E,2021iSci...24j2557C,2021ApJ...907...20K,2025ApJ...990...10S}. Moreover, in some sources, the optical/UV variations can temporarily lead the X-ray emission, indicating that at least part of the disk variability is not simply a passive response to coronal illumination \cite[e.g.,][]{2015ApJ...806..129E,2022MNRAS.512L..33V}. These results suggest that additional disk-side heating or fluctuation mechanisms may operate together with conventional reprocessing.

A related challenge comes from the variability timescale. Inward propagation of accretion-rate fluctuations has long been considered a natural explanation for correlated variability across disk radii \cite[e.g.,][]{1997MNRAS.292..679L,2008MNRAS.387L..41L}. However, for a standard $\alpha$-disk ($\alpha$ is the viscosity parameter), the viscous timescale at optical/UV-emitting radii is generally much longer than the day-to-month variability observed in many AGNs, leading to the so-called quasar viscosity problem \cite[e.g.,][]{2018NatAs...2..102L}. This motivates models in which magnetic processes, rather than purely viscous diffusion, communicate fluctuations through the disk and corona. Related thermal-fluctuation models have also shown that UV/optical inter-band correlations and lags can arise without invoking standard light-echo reprocessing \citep[e.g.,][]{2018ApJ...855..117C}. In the corona-heated accretion-disk reprocessing (CHAR) framework, magnetic fluctuations generated in the corona are coupled to the disk and can reproduce several statistical properties of AGN optical/UV variability, including wavelength-dependent amplitudes, damping timescales, and inter-band lags \cite[e.g.,][]{2020ApJ...891..178S}. Nevertheless, the physical origin of the magnetic fluctuations remains an open issue.

AGN disks are unlikely to be smooth, isolated fluids. They may contain a population of embedded stars and compact objects supplied either by the capture of pre-existing nuclear star cluster members or by in-situ formation in gravitationally unstable outer disk regions \cite[e.g.,][]{2003MNRAS.341..501S,2007MNRAS.374..515L,2016ApJ...819L..17B,2019ApJ...878...85S,2022ApJ...928..191G,2024ApJ...966L...9Z}. Among these objects, stellar-mass black holes (sBHs) are of particular interest because they can accrete from the surrounding gas, exchange angular momentum with the disk, and migrate radially under disk torques. By analogy with planet migration in protoplanetary disks, low-mass embedded objects undergo Type I migration, whereas more massive objects can open gaps and transition toward slower gap-modified or Type II migration \cite[e.g.,][]{1996Icar..124...62P,2002ApJ...565.1257T,2010apf..book.....A,2018ApJ...861..140K}. Because migration torques depend sensitively on the radial structure of the disk, changes in surface density, temperature, and entropy/vortensity can create radii where inward and outward migration balance. These radii act as migration traps, causing embedded objects to converge and accumulate \cite[e.g.,][]{2016ApJ...819L..17B}. Such pile-ups can strongly enhance the local compact-object density, increase interaction rates, and may leave observable imprints on the disk emission \cite[e.g.,][]{2012MNRAS.425..460M,2014MNRAS.441..900M}. Such gas-assisted compact-object populations may also have implications for BH mergers: recent GWTC-3 population analyses have identified a high-mass, high-spin subpopulation consistent with hierarchical mergers, while subsequent work has discussed a possible contribution from AGN-disk environments \cite[e.g.,][]{2024PhRvL.133e1401L,2025ApJ...987...65L}. Recent related studies have further suggested that embedded sBHs may leave observable continuum reverberation signatures in accretion disks \cite[e.g.,][]{2025arXiv251107716W} and that similar disk-embedded compact-object scenarios may be relevant to little red dots at high redshift \cite[e.g.,][]{2025arXiv251109278W}.

Recent hydrodynamic studies further indicate that the migration of accreting embedded objects is not necessarily monotonic inward. When gas accretion onto the embedded object becomes efficient, asymmetric flows in the circum-object region can generate positive torques and drive outward migration over a specific range of gap-opening parameters \cite[e.g.,][]{2018ApJ...861..140K,2015MNRAS.448..994K,2026ApJ...997..160I,2026ApJ...997..161P}. This behavior implies that the migration direction, trap location, and compact-object accumulation efficiency are sensitive to the background AGN disk state, especially to the aspect ratio, surface density, and accretion rate  \cite[e.g.,][]{2012MNRAS.425..460M,2014MNRAS.441..900M,2020ApJ...898...25T,2023MNRAS.526.5346L}. Consequently, embedded sBHs may not merely act as passive tracers of the disk; their spatial distribution can become dynamically concentrated at preferred radii, creating localized regions where additional heating and variability are expected to be strongest \cite[e.g.,][]{2021ApJ...906...52L,2024ApJ...971..130L}.

Magnetic reconnection provides a natural channel for converting the interaction between embedded sBHs and the magnetized disk into radiation. Reconnection rapidly transforms magnetic energy into particle heating and nonthermal emission in a wide variety of astrophysical plasmas \cite[e.g.,][]{2007PhPl...14j0703L,2010PhRvL.105w5002U,2019PhRvL.122e5101Z}. In AGN environments, it has been invoked in connection with coronal heating, jet activity, and high-energy flares \cite[e.g.,][]{2004ApJ...606.1083H,2009MNRAS.395.2183Y,2018ApJ...862L..25Z,2022ApJ...924..124Y,2023A&A...677A..67E,2020JGRA..12525935H}. Particle-in-cell simulations show that relativistic reconnection can reach high radiative efficiencies and generate intermittent, burst-like energy release \cite[e.g.,][]{2021JPlPh..87e9012R,2022ApJ...924L..32R,2018ApJ...852...95N}. Within a magnetically coupled disk-corona system, the accumulation and transport of magnetic flux can therefore play a central role in regulating AGN variability \cite[e.g.,][]{2013ApJ...764L..24S,2013MNRAS.431..355G,2020ApJ...891..178S}.

Motivated by this picture, \cite{2025ApJ...991..167X} proposed that the motion of sBHs through a magnetized AGN disk can trigger magnetic reconnection. In that model, disk-side reconnection first produces a UV/EUV-peaking flare, while heated magnetized plasma may subsequently rise into the corona and trigger delayed X-ray emission \cite[e.g.,][]{2018ApJ...862L..25Z,2022ApJ...924..124Y}. Such a two-stage reconnection scenario may have implications for UV/X-ray lag reversals \cite[e.g.,][]{2021iSci...24j2557C}. In the present work, however, we focus on the disk-side consequences of this mechanism: how a population of migrating and trapped sBHs modifies the optical/UV thermal variability of an AGN disk. The X-ray response associated with secondary coronal reconnection is left as a natural extension and will be analyzed separately.

Here we combine the CHAR disk-variability framework by \cite{2020ApJ...891..178S} with a population model for embedded sBHs undergoing accretion-modified migration in AGN disks \cite[e.g.,][]{2026ApJ...997..160I,2026ApJ...997..161P}. This framework is supported by its successful applications to observed AGN UV/optical variability, structure functions (SFs), and interband continuum lags \cite[e.g.,][]{2020ApJ...891..178S,2020ApJ...902....7S,2024ApJ...962..134C,2024ApJ...974..271L,2024ApJ...966L...9Z}. We semi-analytically evolve the radial and mass distribution of sBHs under different SMBH accretion states \cite[e.g.,][]{2003MNRAS.341..501S} and compute the magnetic reconnection heating induced by their interaction with the local disk gas and magnetic field \cite[e.g.,][]{2025ApJ...991..167X}. The resulting heating is implemented as a stochastic, spatially localized shot-noise process superposed on the background magnetic disk fluctuations \cite[e.g.,][]{2005MNRAS.359..345U}. This allows us to synthesize optical/UV light curves and quantify how embedded sBHs affect the thermal response, time-domain and lag diagnostics, SFs, and inter-band lag-wavelength relation.

The central question addressed in this paper is whether migration traps can act as state-dependent variability attractors in AGN disks, and whether the resulting concentration of embedded sBHs can leave detectable signatures in the optical/UV variability of AGNs. We address this question by tracking the coupled evolution of the embedded-sBH population, the associated magnetic reconnection heating, and the resulting disk variability across a range of SMBH accretion states.

The paper is organized as follows. In Section~\ref{sec:model}, we describe the background AGN disk and CHAR-like thermal variability, the embedded sBH population synthesis, the accretion-modified migration prescription, and the magnetic reconnection heating model. In Section~\ref{sec:results}, we present the migration, the radial heating structure, and the optical/UV variability results. Section~\ref{sec:discussion and conclusion} discusses the observational implications, limitations of the current model, the connection to future X-ray extensions, and summarizes our conclusions.

\section{Model and Methods}
\label{sec:model}

\subsection{Background Disk and CHAR-like Magnetic Heating}
\label{subsec:disk}

Following \citet{1973A&A....24..337S}, we adopt a geometrically thin, optically thick standard $\alpha$-disk as the background AGN disk. The central SMBH has mass $M_{\bullet}$, dimensionless accretion rate $\dot{m}=\dot{M}_{\bullet}/\dot{M}_{\rm Edd,\bullet}$, where $\dot{M}_{\rm Edd,\bullet}$ is the Eddington accretion rate of the SMBH. The radial computational domain extends from $R_{\rm in}=10R_{\rm S}$ to $R_{\rm out}=5000R_{\rm S}$ with logarithmic spacing, where $R_{\rm S}=2GM_{\bullet}/c^2$ is the Schwarzschild radius. Throughout this work, the disk structure is first constructed as a steady-state background, which later serves as the environment for sBH migration, magnetic reconnection, and optical/UV light-curve synthesis.

For a Keplerian thin disk, the angular frequency is $\Omega_{\rm K}=({GM_{\bullet}}/{R^3})^{1/2}$ \cite[e.g.,][]{1973A&A....24..337S,2008bhad.book.....K,2020ApJ...891..178S}. The vertical structure is described by hydrostatic balance, $H={c_s}/{\Omega_{\rm K}}$, where $H$ is the disk scale height and $c_s$ is the sound speed \cite[e.g.,][]{2008bhad.book.....K,2020ApJ...891..178S}. The unperturbed mid-plane density is then approximated as $\rho_0(R)={\Sigma(R)}/{2H(R)}$, where $\Sigma(R)$ is the surface density. These quantities provide the local gas environment used in the subsequent calculations of embedded-sBH migration, gap opening, ram pressure, and magnetic reconnection. Unless otherwise stated, we adopt the fiducial parameters listed in Table~\ref{tab:parameters}.

The local viscous heating rate per unit disk surface is written as
\begin{equation}
    Q_{\rm vis}^0(R)
    =
    \frac{3GM_{\bullet}\dot{M}_{\bullet}}{8\pi R^3}
    \left[1-\left(\frac{R_{\rm in}}{R}\right)^{1/2}\right],
    \label{eq1}
\end{equation}
following the standard thin-disk solution \cite[e.g.,][]{1973A&A....24..337S,2020ApJ...891..178S}. Here, the superscript ``0'' denotes the stationary background value before time-dependent fluctuations and sBH-driven heating are applied. The steady disk temperature and surface density are obtained from the usual vertical energy balance,
\begin{equation}
    \frac{8\sigma_{\rm SB}T_c^4}{3\tau_\mathrm{d}}
    =
    Q_{\rm vis}^0(R)+Q_{\rm mc}^0(R),
    \label{eq2}
\end{equation}
together with the equation of state and opacity prescription \cite[e.g.,][]{2008bhad.book.....K,2020ApJ...891..178S}. Here $Q_{\rm mc}^0$ is the CHAR magnetic corona heating term, $T_c$ is the mid-plane temperature, $\tau_\mathrm{d}$ is the optical depth, and $\sigma_{\rm SB}$ is the Stefan-Boltzmann constant.

Following the CHAR framework, we include an additional magnetic-corona heating component that represents energy transported from the magnetized corona into the disk \cite[e.g.,][]{2013ApJ...764L..24S,2020ApJ...891..178S}. In the stationary background model, this component is parameterized as a fixed fraction of the local viscous heating, $Q_{\rm mc}^0(R)=k_{\rm ratio}Q_{\rm vis}^0(R)$, where we adopt $k_{\rm ratio}=1/3$ as the fiducial value, following the fiducial CHAR setup of \cite{2020ApJ...891..178S}. Thus, the total stationary heating that determines the initial disk structure is $Q_{\rm disk}^0(R)=Q_{\rm vis}^0(R)+Q_{\rm mc}^0(R)$.

To model intrinsic disk variability, both the viscous and magnetic-corona heating components are later modulated by a common stochastic magnetic-fluctuation field. 
Specifically, we write
\begin{equation}
    Q_{\rm vis}^+(R,t)
    =
    Q_{\rm vis}^0(R) f_{\rm mc}(R,t),
    \label{eq3}
\end{equation}
and
\begin{equation}
    Q_{\rm mc}^+(R,t)
    =
    Q_{\rm mc}^0(R) f_{\rm mc}(R,t),
    \label{eq4}
\end{equation}
where $f_{\rm mc}(R,t)$ is a dimensionless, mean-normalized lognormal red-noise process. This prescription follows the CHAR assumption that coronal magnetic fluctuations are coupled to the disk and drive coherent disk-temperature fluctuations \cite[e.g.,][]{2020ApJ...891..178S}. The adopted red-noise behavior is also motivated by propagating accretion-fluctuation models and coronal magnetohydrodynamics (MHD) dissipation with approximately $1/f$-like power spectra, while the lognormal distribution follows the commonly observed flux variability statistics of accreting systems \cite[e.g.,][]{1997MNRAS.292..679L,2004MNRAS.348..111K,2009ApJ...703..964N,2005MNRAS.359..345U,2020ApJ...891..178S}. This prescription captures the background magnetic variability of the disk-corona system, while the additional heating induced by embedded sBHs is introduced separately in Subsection~\ref{subsec:sbh_heating}.

\subsection{Embedded sBH Population and Migration-Trap Formation}
\label{subsec:population}

We evolve the embedded sBH population in the two-dimensional $(R,M)$ state space, where $R$ is the cylindrical radius and $M$ is the instantaneous sBH mass. The distribution function $f(R,M,t)$ is defined such that $dN=2\pi Rf\,dMdR$ gives the expected number of the embedded sBHs in the mass interval $(M,M+dM)$ and annulus $(R,R+dR)$. Conservation of object number then gives the population-balance equation,
\begin{equation}
\frac{\partial f}{\partial t}+\frac{1}{R}\frac{\partial}{\partial R}\left(R v_{\mathrm{R}} f\right)+\frac{\partial}{\partial M}\left(\dot{M}_{\mathrm{sBH}} f\right)=S(R,M,t),
\label{eq5}
\end{equation}
where $v_{\mathrm{R}}(R,M)=dR/dt$ is the radial migration velocity, $\dot{M}_{\mathrm{sBH}}(R,M)=dM/dt$ is the gas accretion rate onto the sBH. The first advection term describes transport of sBH number through cylindrical annuli, with the factor $1/R$ arising solely from the annular geometry, while the second describes transport between mass bins caused by accretion. No incompressibility condition is assumed. The source term $S(R,M,t)$ represents continuous sBH formation and capture. In the present model, we neglect explicit diffusion or collisional redistribution in $(R,M)$ due to stochastic migration, gravitational scattering, binary interactions, mergers, or ejections. Similar number-conservation formulations are commonly used for accretion-driven evolution of BH mass distributions \cite[e.g.,][]{1992MNRAS.259..725S,2002MNRAS.335..965Y}, while extensions with mass as an additional state-space coordinate have been considered for accreting stellar and sBH populations \cite[e.g.,][]{2023MNRAS.520.4502W}. With this definition, $f$ has units of $\mathrm{cm^{-2}\,g^{-1}}$, while the source term has units of $\mathrm{cm^{-2}\,g^{-1}\,s^{-1}}$.

\subsubsection{Migration Physics}
\label{subsubsec:migration}

We calculate the radial migration of embedded sBHs using the semi-analytical framework motivated by recent hydrodynamic studies of accreting embedded objects \cite[e.g.,][]{2024ApJ...971..130L,2024ApJ...975..296L,2026ApJ...997..160I,2026ApJ...997..161P}. For low-mass objects that do not open a deep gap, the baseline Type I migration velocity is written as
\begin{equation}
v_{\mathrm{I,base}}
=
-
C_{\mathrm{I}}
q
\left(
\frac{\Sigma R^2}{M_{\bullet}}
\right)
h^{-2}
R\Omega_{\mathrm{K}},
\label{eq6}
\end{equation}
where $q={M}/{M_{\bullet}}$is the sBH-to-SMBH mass ratio. The factor $\Sigma R^2/M_{\bullet}$ is dimensionless, $q$ and $h=H/R$ are also dimensionless, and $R\Omega_{\mathrm{K}}$ has units of velocity; therefore Equation~\eqref{eq6} has the correct dimensions. For a locally isothermal disk with a typical surface density profile close to $\Sigma\propto R^{-1}$, we adopt $C_{\mathrm{I}}=3.8$ \cite[e.g.,][]{2002ApJ...565.1257T,2026ApJ...997..160I}.

As the sBH grows, its tidal torque partially depletes the gas near its orbit and slows the migration. We describe this effect using the gap-opening parameter calibrated by hydrodynamic gap models \cite[e.g.,][]{2015MNRAS.448..994K,2018ApJ...861..140K} 
\begin{equation}
K'
=
0.04\,
\frac{q^2}{h^5\alpha},
\label{eq7}
\end{equation}
which is dimensionless. The corresponding gap-reduced migration velocity is
\begin{equation}
v_{\mathrm{gap}}
=
f_{\mathrm{gap}}v_{\mathrm{I,base}}
=
\frac{v_{\mathrm{I,base}}}{1+K'}.
\label{eq8}
\end{equation}
This form follows the interpretation that gap-opening objects migrate approximately as Type-I perturbers embedded in the depleted gas at the gap bottom \cite[e.g.,][]{2018ApJ...861..140K}.

Gas accretion onto the sBH can further reduce the background mass flux that passes through the orbit \cite[e.g.,][]{2016ApJ...823...48T,2020MNRAS.498.2054R,2020ApJ...891..143T}. We account for this effect by defining $v_{\mathrm{dep}}=f_{\mathrm{pass}}v_{\mathrm{gap}}$, where $f_{\mathrm{pass}}$ is the fraction of the unperturbed disk mass flux that bypasses the sBH without being accreted. The background disk mass flux is $\dot{M}_{\mathrm{disk}}=3\pi\nu\Sigma$, where $\nu=\alpha c_{\mathrm{s}}H$ is the effective kinematic viscosity. The bypass fraction is computed semi-analytically from the sBH accretion rate in Section~\ref{subsubsec:mass_accretion}.

Following the semi-analytical formula of \cite{2026ApJ...997..160I}, calibrated against the parameter survey of \cite{2026ApJ...997..161P}, the migration velocity changes continuously across inward, outward, and deep-gap inward regimes. The transition boundaries are controlled by the local aspect ratio $h$. We define $K'_{\mathrm{acc}}$ at which the accretion-driven positive torque starts to cancel inward migration, and $K'_{\mathrm{acc,l}}$ as the lower edge of the transition region, i.e.,
\begin{equation}
K'_{\mathrm{acc}}
=
0.01\left(\frac{h}{0.05}\right),
\qquad
K'_{\mathrm{acc,l}}
=
K'_{\mathrm{acc}}10^{-0.5},
\label{eq9}
\end{equation}
and adopt $K'_{\mathrm{gap}}=50$, $K'_{\mathrm{gap,h}}=250$ to represent the critical points for initiating shallow grooves and deep gaps. The full migration velocity is then given by the following piecewise prescription.

\begin{enumerate}
\item For shallow gaps, $K'\leq K'_{\mathrm{acc,l}}$, the object undergoes inward Type I-like migration modified by gap opening and gas depletion: $v_{\mathrm{R}}=v_{\mathrm{dep}}$.

\item For the first transition region, $K'_{\mathrm{acc,l}}<K'<K'_{\mathrm{acc}}$, we interpolate logarithmically from inward migration to the zero-velocity boundary:
\begin{equation}
v_{\mathrm{R}}
=
v_{\mathrm{dep}}
\left[
1-
\frac{
\log\left(K'/K'_{\mathrm{acc,l}}\right)
}{
\log\left(K'_{\mathrm{acc}}/K'_{\mathrm{acc,l}}\right)
}
\right].
\label{eq10}
\end{equation}
Thus $v_{\mathrm{R}}=0$ at $K'=K'_{\mathrm{acc}}$, which defines the inner null-velocity boundary of the migration trap.

\item For intermediate gap depths, $K'_{\mathrm{acc}}\leq K'\leq K'_{\mathrm{gap}}$, efficient gas accretion can generate an asymmetric horseshoe flow and a positive corotation torque \cite[e.g.,][]{2024ApJ...971..130L,2024ApJ...975..296L,2026ApJ...997..160I,2026ApJ...997..161P}. The radial velocity is prescribed as
\begin{equation}
v_{\mathrm{R}}
=
-
v_{\mathrm{dep}}
\frac{
\log\left(K'/K'_{\mathrm{acc}}\right)
}{
\log\left(K'_{\mathrm{gap}}/K'_{\mathrm{acc}}\right)
}.
\label{eq11}
\end{equation}
Because $v_{\mathrm{dep}}<0$ for inward Type I migration, Equation~\eqref{eq11} gives $v_{\mathrm{R}}>0$, corresponding to outward migration.

\item For deeper gaps, $K'_{\mathrm{gap}}<K'<K'_{\mathrm{gap,h}}$, the outward torque is gradually suppressed. We again use a logarithmic interpolation,
\begin{equation}
\begin{split}
v_{\mathrm{R}}
=
v_{\mathrm{out,gap}}
&+
\left(
v_{\mathrm{in,gap,h}}
-
v_{\mathrm{out,gap}}
\right)
\\
&\quad\times
\frac{
\log\left(K'/K'_{\mathrm{gap}}\right)
}{
\log\left(K'_{\mathrm{gap,h}}/K'_{\mathrm{gap}}\right)
},
\end{split}
\label{eq12}
\end{equation}
where $v_{\mathrm{out,gap}}=-v_{\mathrm{dep}}(K'_{\mathrm{gap}})$ is $v_{\mathrm{dep}}$ evaluated at the local point satisfying $K'=K'_{\mathrm{gap}}$, and $v_{\mathrm{in,gap,h}}=v_{\mathrm{gap}}(K'_{\mathrm{gap,h}})$ is $v_{\mathrm{gap}}$ evaluated at $K'=K'_{\mathrm{gap,h}}$. The outer migration trap is located within this transition region, where the interpolated $v_{\mathrm{R}}$ crosses zero.

\item Finally, for very deep gaps, $K'\geq K'_{\mathrm{gap,h}}$, the positive accretion-driven torque is quenched, and the object migrates inward with the gap-modified velocity, $v_{\mathrm{R}}=v_{\mathrm{gap}}$. For numerical stability, we impose an upper limit on the absolute radial migration speed, $|v_{\mathrm{R}}|\leq0.1R\Omega_{\mathrm{K}}$. This cap only prevents unphysically large velocities in poorly resolved transition regions and does not affect the location of the migration traps.
\end{enumerate}

Stable migration traps form at the null-velocity points where $v_{\mathrm{R}}=0$ and the migration direction reverses. In the present prescription, the inner trap is located at $K'=K'_{\mathrm{acc}}$, while the outer trap appears inside the second transition region. Because $K'$ depends on $q$, $h$, and $\alpha$, and because $h(R)$ is set by the background AGN disk structure, the trap locations are anchored to the thermodynamic state of the disk. These traps, therefore, act as phase-space attractors that collect sBHs over long-term population evolution.

\subsubsection{Gas Accretion onto Embedded sBHs}
\label{subsubsec:mass_accretion}

The gas accretion rate onto each sBH determines both its mass growth and the bypass fraction $f_{\mathrm{pass}}$ entering the migration velocity. We compute $\dot{M}_{\mathrm{sBH}}$ using the thermal mass ratio $q_{\mathrm{th}}={q}/{h^3}$, which measures the Hill radius relative to the disk scale height \cite[e.g.,][]{2023MNRAS.525.2806C,2023MNRAS.526.5346L,2026ApJ...997..160I,2026ApJ...997..161P}. As the sBH opens a gap, the effective surface density available for accretion is reduced to $\Sigma_{\mathrm{local}}={\Sigma}/({1+K'})$.

For low thermal masses, $q_{\mathrm{th}}\lesssim 1$, the local kinematic accretion rate is approximated by the Bondi-like scaling \cite[e.g.,][]{1952MNRAS.112..195B,2023MNRAS.525.2806C,2026ApJ...997..160I}
\begin{equation}
\dot{M}_{\mathrm{B}}
=
\left(\frac{\pi}{2}\right)^{1/2}
q_{\mathrm{th}}^2
\Sigma_{\mathrm{local}}
(hR)^2
\Omega_{\mathrm{K}}.
\label{eq13}
\end{equation}
For high thermal masses, $q_{\mathrm{th}}\gtrsim 1$, the accretion flow becomes controlled by the Hill scale. We use the Hill-like scaling motivated by simulations of accretion onto embedded gap-opening objects \cite[e.g.,][]{2023MNRAS.525.2806C,2023MNRAS.526.5346L,2026ApJ...997..161P}
\begin{equation}
\dot{M}_{\mathrm{H}}
=
3
\left(
\frac{q_{\mathrm{th}}}{3}
\right)^{2/3}
\Sigma_{\mathrm{local}}
(hR)^2
\Omega_{\mathrm{K}}.
\label{eq14}
\end{equation}
To connect the two limits smoothly, we define the local kinematic accretion demand as $\dot{M}_{\mathrm{kin}}=w_{\mathrm{B}}\dot{M}_{\mathrm{B}}+w_{\mathrm{H}}\dot{M}_{\mathrm{H}}$, with
\begin{equation}
w_{\mathrm{B}}
=
\frac{1}{1+q_{\mathrm{th}}^2},
\qquad
w_{\mathrm{H}}
=
\frac{q_{\mathrm{th}}^2}{1+q_{\mathrm{th}}^2}.
\label{eq15}
\end{equation}
We then impose an Eddington cap on the local accretion demand,
\begin{equation}
\dot{M}_{\mathrm{demand}}
=
\min
\left(
\dot{M}_{\mathrm{kin}},
\dot{M}_{\mathrm{Edd,sBH}}
\right),
\label{eq16}
\end{equation}
where $\dot{M}_{\mathrm{Edd,sBH}}=L_{\mathrm{Edd,sBH}}/(\eta c^2)$ is the Eddington accretion rate of the sBH, $L_{\mathrm{Edd,sBH}}$ is its Eddington luminosity, and the radiative efficiency is $\eta=0.1$.

The actual sBH growth rate cannot exceed the mass flux supplied by the background AGN disk. Following the supply-limited accretion treatment adopted for accreting gap-opening objects \cite[e.g.,][]{2016ApJ...823...48T,2020MNRAS.498.2054R,2026ApJ...997..160I}, we define the accretion demand ratio as $\xi={\dot{M}_{\mathrm{demand}}}/{\dot{M}_{\mathrm{disk}}}$, and write the fraction of the background disk flux captured by the sBH as $f_{\mathrm{p}}={\xi}/({1+\xi})$. The final accretion rate is $\dot{M}_{\mathrm{sBH}}=f_{\mathrm{p}}\dot{M}_{\mathrm{disk}}$. The remaining fraction of the gas flux bypasses the sBH orbit,
\begin{equation}
f_{\mathrm{pass}}
=
1-f_{\mathrm{p}}
=
\frac{1}{1+\xi},
\label{eq17}
\end{equation}
and feeds back into the migration velocity. This coupling ensures that gap opening, gas accretion, and migration are treated consistently within the one-dimensional population synthesis.

\subsubsection{Initial Conditions and Continuous Replenishment}
\label{subsubsec:initial_conditions}

The embedded sBH population is supplied by two physically motivated channels. The first channel is the capture of pre-existing compact objects from the nuclear star cluster (NSC), which can be ground down into the AGN disk through repeated disk crossings and gas drag \cite[e.g.,][]{2017ApJ...835..165B,2019ApJ...878...85S}. The second channel is in-situ formation in the outer disk, where the AGN disk becomes gravitationally unstable and fragments into massive stars that subsequently leave compact remnants \cite[e.g.,][]{2003MNRAS.341..501S,2007MNRAS.374..515L,2022ApJ...928..191G,2026ApJ...999...55C}. Because we focus on continuously replenished embedded populations, the fiducial calculations start with no pre-existing embedded sBHs, i.e. $N_{\rm ref}=0$. The formal initial distribution, therefore, satisfies
\begin{equation}
     \int_{R_\mathrm{in}}^{R_\mathrm{out}} 2\pi R\,dR\int_{M_\mathrm{min}}^{M_\mathrm{max}} dM\,f(R,M,t=0)=N_{\rm ref}=0 . 
     \label{eq18}
\end{equation} 
The subsequent embedded population is therefore supplied entirely by the continuous source terms described below.

The NSC-capture component is described by a centrally concentrated radial source template,
\begin{equation}
\Sigma_{\mathrm{cap}}(R)
\propto
R^{-\gamma_{\mathrm{cap}}},
\qquad
R\geq R_{\mathrm{in,cut}},
\label{eq19}
\end{equation}
where we take the capture profile index $\gamma_{\mathrm{cap}}=1.5$. An inner cutoff $R_{\mathrm{in,cut}}=10R_{\mathrm{S}}$ is imposed to avoid an artificial divergence near the inner disk boundary. This power-law form approximates the projected distribution expected for compact objects supplied from the inner nuclear cluster \cite[e.g.,][]{2017ApJ...835..165B,2018MNRAS.478.4030G,2019ApJ...878...85S}.

The in-situ component is localized near the radius where the disk first becomes susceptible to gravitational fragmentation. We identify this radius, $R_{\mathrm{insitu}}$, from the background disk structure by requiring that the Toomre parameter $Q_{\mathrm{T}}={c_{\mathrm{s}}\Omega_{\mathrm{K}}}/({\pi G\Sigma})$ approaches unity \cite[e.g.,][]{2001ApJ...553..174G,2003MNRAS.339..937G}. Here $Q_{\mathrm{T}}$ is dimensionless, since $c_{\mathrm{s}}\Omega_{\mathrm{K}}$ and $\pi G\Sigma$ both have units of acceleration. The radial distribution of the in-situ component is then modeled as a narrow log-normal ring,
\begin{equation}
\Sigma_{\mathrm{insitu}}(R)
\propto
\exp
\left[
-\frac{1}{2}
\left(
\frac{\ln(R/R_{\mathrm{insitu}})}{\sigma_{\mathrm{insitu}}}
\right)^2
\right],
\label{eq20}
\end{equation}
where we adopt $\sigma_{\mathrm{insitu}}=0.15$ as the logarithmic width of the radial distribution for the in-situ formed sBH population. This localized profile reflects the fact that disk fragmentation is expected to occur only over a limited radial range in AGN disks \cite[e.g.,][]{2007MNRAS.374..515L,2022ApJ...928..191G}.

The in-situ source mass distribution is assumed to be top-heavy,
\begin{equation}
\frac{dN_\mathrm{insitu}}{dM}
\propto
M^{-\alpha_{\mathrm{M}}},
\qquad
M_{\mathrm{min}}\leq M\leq M_{\mathrm{max}},
\label{eq21}
\end{equation}
with the in-situ mass function slope $\alpha_{\mathrm{M}}=1.3$, $M_{\mathrm{min}}=8M_{\odot}$, and $M_{\mathrm{max}}=1000M_{\odot}$. This choice is motivated by expectations for star formation and compact-remnant production in dense AGN disks and other extreme environments \cite[e.g.,][]{2017MNRAS.464..946S,2022ApJ...928..191G}. 

Continuous replenishment is included through the source term in Equation~\eqref{eq5}. We write the source as the sum of capture and in-situ contributions,
\begin{equation}
S(R,M,t)
=S_{\mathrm{cap}}(R,M,t)+S_{\mathrm{insitu}}(R,M,t).
\label{eq22}
\end{equation}
The in-situ component is normalized by an effective star-formation rate $\dot{M}_{\mathrm{sf}}=0.05\,M_{\odot}\,\mathrm{yr}^{-1}$. For an sBH formation efficiency $\eta_{\mathrm{BH}}=0.03 ~M_\odot^{-1}$, this corresponds to an in-situ sBH injection rate $\dot{N}_{\mathrm{insitu}}\sim1.5\times10^{-3}\,\mathrm{yr}^{-1}$ \cite[e.g.,][]{2017MNRAS.464..946S}. The capture component is normalized independently by the adopted capture rate $\dot N_{\rm cap}=10^3\,{\rm Myr}^{-1}$. The in-situ source uses the fiducial top-heavy mass function with $\alpha_M=1.3$, whereas the captured component is assigned a separate top-heavy mass function with ${dN_\mathrm{cap}}/{dM}\propto M^{-\alpha_{\mathrm{cap}}}$, where $\alpha_{\mathrm{cap}}=1.0$. In this form, $S(R,M,t)$ has units of $\mathrm{cm^{-2}\,g^{-1}\,s^{-1}}$, consistent with the population-balance equation. The radial templates in Equations~(\ref{eq19})-(\ref{eq20}) are normalized over the disk surface so that their integrals reproduce the adopted total injection rates, $\dot N_{\rm cap}$ and $\dot N_{\rm insitu}$, respectively. The mass dependence is assigned through the corresponding top-heavy mass functions.

Solving Equation~\eqref{eq5} with the migration and accretion prescriptions described above yields the time-dependent sBH distribution in the disk. For use in the magnetic heating calculation, we reduce the two-dimensional distribution to the radial number and mass distributions,
\begin{equation}
N_{\mathrm{R}}(R,t)
=
2\pi R
\int_{M_{\mathrm{min}}}^{M_{\mathrm{max}}} f(R,M,t)\,dM,
\label{eq23}
\end{equation}
and
\begin{equation}
M_{\mathrm{R}}(R,t)
=
2\pi R
\int_{M_{\mathrm{min}}}^{M_{\mathrm{max}}} M f(R,M,t)\,dM.
\label{eq24}
\end{equation}
Here $N_{\mathrm{R}}(R,t)\,dR$ is the number of sBHs in the annulus $(R,R+dR)$, while $M_{\mathrm{R}}(R,t)\,dR$ is their total mass. The local mean sBH mass is therefore $\langle M_{\mathrm{sBH}}\rangle(R,t)={M_{\mathrm{R}}(R,t)}/{N_{\mathrm{R}}(R,t)}$. These quantities provide the direct bridge to the next section: $N_{\mathrm{R}}(R,t)$ controls the rate of sBH-triggered reconnection events, while $\langle M_{\mathrm{sBH}}\rangle(R,t)$ determines the local gap-opening strength, ram pressure, and characteristic magnetic energy released by each event. The corresponding surface number and mass densities are $\Sigma_N=N_R/(2\pi R)$ and $\Sigma_M=M_R/(2\pi R)$, respectively. All spatial and mass distributions are normalized so that the initial population contains ($N_{\mathrm{ref}}=0$) sBHs and the source terms reproduce the adopted capture and formation rates.

\subsection{sBH-driven Magnetic Heating of the Disk}
\label{subsec:sbh_heating}

We now convert the time-dependent embedded-sBH distribution into an additional magnetic heating source for the disk. The model follows the sBH-triggered reconnection picture proposed by \cite{2025ApJ...991..167X}: as embedded sBHs move through the magnetized AGN disk, shear and turbulent motions stretch the surrounding magnetic field lines and can trigger reconnection near the compact object. This picture is motivated by the general expectation that current sheets and plasmoid-mediated reconnection are common in magnetized accretion flows \cite[e.g.,][]{2007PhPl...14j0703L,2010PhRvL.105w5002U,2020MNRAS.495.1549N,2021JPlPh..87e9012R,2022ApJ...924L..32R}. In this work we only model the disk-side heating that affects the optical/UV emission, while the possible delayed coronal and X-ray response is left for future study.

The population synthesis provides $N_{\mathrm{R}}(R,t)$ and $\langle M_{\mathrm{sBH}}\rangle(R,t)$, which determine the number of reconnection sites and the characteristic sBH mass at each radius. We construct the one-dimensional gap attenuation factor by evaluating the gap-opening correction at the local mean sBH mass,
\begin{equation}
f_{\mathrm{gap}}^{\mathrm{1D}}(R,t)
=
\max\left[
f_{\mathrm{gap}}\left(R,\langle M_{\mathrm{sBH}}\rangle\right),
f_{\mathrm{gap,min}}
\right],
\label{eq25}
\end{equation}
where $f_{\mathrm{gap}}(R,\langle M_{\mathrm{sBH}}\rangle)$ is evaluated from the gap attenuation $1/(1+K')$, and $f_{\mathrm{gap,min}}=0.5$ is an imposed floor on the gap depth. The gap-depth scaling $1/(1+K')$ is calibrated from two-dimensional, midplane hydrodynamic gap models \cite[e.g.,][]{2015MNRAS.448..994K,2018ApJ...861..140K}, which do not capture the vertical gas circulation that replenishes the gap in three dimensions. Three-dimensional simulations of gap-opening embedded objects show that meridional flows feed gas back into the gap from above and below the midplane, so that the true gap floor is substantially shallower than the extrapolation of the two-dimensional scaling into the deep-gap regime $K'\gg1$ \cite[e.g.,][]{2015MNRAS.448..994K}. Because our one-dimensional population synthesis evolves only the radial distribution and does not resolve this vertical replenishment, applying the unmodified $1/(1+K')$ scaling would overestimate the local gas depletion at the deepest gaps. We therefore cap the depletion at $f_{\mathrm{gap,min}}=0.5$, corresponding to at most a $50\%$ reduction in the local gas density available for reconnection. It mainly regulates the saturation level of the heating suppression in the most strongly depleted regions.  In annuli where $N_R=0$, we set $f_{\rm gap}^{\rm 1D}=1$ and $Q_{\rm sBH}^{+}=0$; $\langle M_{\rm sBH}\rangle=M_R/N_R$ is evaluated only for annuli with $N_R>0$. The local gap correction $f_{\mathrm{gap}}^{\mathrm{1D}}(R,t)$ is then used to reduce the gas density participating in the reconnection process,
\begin{equation}
\rho_{\mathrm{c}}^{0}(R,t)
=
f_{\mathrm{gap}}^{\mathrm{1D}}(R,t)\rho_{0}(R).
\label{eq26}
\end{equation}
This is the first role of gap opening in our heating model: a smaller $f_{\mathrm{gap}}^{\mathrm{1D}}$ implies a lower co-moving gas density, weaker ram pressure, and therefore less efficient reconnection. Thus, sBH pile-up at a migration trap does not necessarily produce a monotonic increase in heating; once the accumulated objects open deep gaps, the local gas supply can become depleted and the reconnection power can be partially choked.

\subsubsection{Reconnection Triggering}
\label{subsubsec:reconnection_trigger}

The magnetic pressure associated with the background CHAR-like magnetic heating is estimated as $P_{0}(R)={Q_{\mathrm{mc}}^{0}(R)}/({2H\Omega_{\mathrm{K}}})$ \cite[e.g.,][]{2020ApJ...891..178S}, with $B_{0}=(8\pi P_{0})^{1/2}$. The ram pressure of the gas interacting with the sBH is
\begin{equation}
P_{\mathrm{r}}(R,t)
=
\rho_{\mathrm{c}}(R,t)v_{\mathrm{rel}}^{2}(R,t).
\label{eq27}
\end{equation}
The use of ram pressure as the triggering agent follows the basic picture that sufficiently strong plasma motion can distort magnetic-field topology and initiate reconnection \cite[e.g.,][]{2017ApJ...836L..32Z,2020JGRA..12525935H,2025ApJ...991..167X}. For Type-I-like objects, we include the density modification induced by the horseshoe flow,
\begin{equation}
\rho_{\mathrm{c}}
=
\rho_{\mathrm{c}}^{0}
(1-\frac{2\zeta}{\hat{\gamma}}
{x_{\mathrm{s}}})
,
\label{eq28}
\end{equation}
where $\zeta=1$ is the power-law index for the initial entropy curve (i.e., $S\propto R^{-\zeta}$, with $\zeta=1$), $\hat{\gamma}=5/3$ is the adiabatic index, and $x_{\mathrm{s}}\approx 1.2\sqrt{q/h}$ is the dimensionless horseshoe half-width in units of $R$. We adopt $\rho_{\mathrm{c}}=\rho_{\mathrm{c}}^{0}$ for non-Type-I objects \cite[e.g.,][]{2009MNRAS.394.2297P,2024ApJ...971..130L}. The relative velocity is taken to be $v_{\mathrm{rel}}=\left(v_{\mathrm{shear}}^{2}+v_{\mathrm{turb}}^{2}\right)^{1/2}$, where the shear velocity is defined as $v_{\mathrm{shear}}\simeq(3/2)\Omega_{\mathrm{K}}R_{\mathrm{Hill}}$, with the Hill radius $R_{\mathrm{Hill}}=R(\langle M_{\mathrm{sBH}}\rangle/3M_{\bullet})^{1/3}$, and the turbulent velocity is given by $v_{\mathrm{turb}}\simeq\alpha c_{\mathrm{s}}$ \cite[e.g.,][]{2020ApJ...898...25T,2021ApJ...906...52L}. This choice is important at migration traps: even when the secular radial drift vanishes, shear and turbulence remain finite and can continue to drive reconnection.

We use a smooth triggering function,
\begin{equation}
\mathcal{G}(R,t)
=
\min
\left[
\frac{P_{\mathrm{r}}(R,t)}{P_{0}(R)},
1
\right],
\label{eq29}
\end{equation}
rather than a hard threshold. Thus, reconnection is inefficient when $P_{\mathrm{r}}\ll P_0$, and saturates when the ram pressure becomes comparable to the magnetic pressure. This prescription represents the intermittent nature of reconnection in turbulent magnetized plasma \cite[e.g.,][]{2019PhRvL.122e5101Z,2023ApJ...946L...3T}.

\subsubsection{Reconnection Energetics}
\label{subsubsec:reconnection_energetics}

Once reconnection is triggered, the magnetic pressure in the current sheet is assumed to scale with the ram pressure, $P_{\mathrm{m}}={P_{\mathrm{r}}}/{\beta_{\mathrm{p}}}$, where $\beta_{\mathrm{p}}=0.01$ is the effective ram-to-magnetic pressure ratio. Low-$\beta_{\mathrm{p}}$ current sheets are motivated by strongly magnetized disk-surface and coronal environments, where magnetic pressure can dominate the local plasma pressure and facilitate fast reconnection \cite[e.g.,][]{2018ApJ...852...95N,2022MNRAS.513.4267N,2025ApJ...991..167X}. The current sheet has length $l_{\mathrm{sh}}=20R_{\mathrm{S,sBH}}$ and width $\delta_{\mathrm{sh}}=g_{{s}}l_{\mathrm{sh}}$, where $R_{\mathrm{S,sBH}}=2G\langle M_{\mathrm{sBH}}\rangle/c^2$. These current-sheet scales are chosen to be consistent with plasmoid-mediated reconnection structures seen in relativistic accretion-flow simulations \cite[e.g.,][]{2021JPlPh..87e9012R,2022ApJ...924L..32R,2025ApJ...991..167X}. The corresponding field-amplification timescale is
\begin{equation}
t_{\mathrm{c}}
=
\frac{
B_{\mathrm{m}}l_{\mathrm{sh}}\delta_{\mathrm{sh}}
}{
B_{0}v_{\mathrm{rel}}(2H)
},
\label{eq30}
\end{equation}
where $B_{\mathrm{m}}=(8\pi P_{\mathrm{m}})^{1/2}$ is magnetic field strength in the current sheet. During one vertical Alfv\'en crossing time, $t_{\mathrm{dyn}}=H/v_{\mathrm{A}}$, the number of magnetic islands is approximated as $N_{\mathrm{mag}}={t_{\mathrm{dyn}}}/{t_{\mathrm{c}}}$, where the Alfv\'en speed $v_{\mathrm{A}}=\sqrt{{\sigma}/({1+\sigma})}c$, and $\sigma=15$ is treated as a prescribed effective magnetization of the reconnecting plasma, rather than being solved self-consistently from the local disk density and magnetic field \cite[e.g.,][]{2013MNRAS.431..355G,2020MNRAS.495.1549N,2021JPlPh..87e9012R}. This estimate follows the idea that multiple current sheets or plasmoids can be generated during one dynamical episode of magnetic compression \cite[e.g.,][]{2007PhPl...14j0703L,2010PhRvL.105w5002U,2025ApJ...991..167X}. The peak heating flux from a single event, averaged over an annulus of width $\delta R$, is
\begin{equation}
Q_{\mathrm{event}}^{\mathrm{peak}}
=
\frac{
N_{\mathrm{mag}}P_{\mathrm{m}}\Omega_{\mathrm{K}}R
l_{\mathrm{sh}}\delta_{\mathrm{sh}}
}{
2\pi R\,\delta R
}.
\label{eq31}
\end{equation}
This expression links the microscopic reconnection event to the macroscopic disk heating rate. Because $P_{\mathrm{m}}\propto P_{\mathrm{r}}\propto\rho_{\mathrm{c}}$, gap depletion directly suppresses the peak event power.

\subsubsection{Mean and Stochastic sBH Heating}
\label{subsubsec:mean_sbh_heating}

The reconnection event rate in an annulus is
\begin{equation}
\dot{\mathcal{N}}_{\mathrm{rec}}
=
\frac{
N_{\mathrm{R}}(R,t)\delta R
}{
t_{\mathrm{dyn}}
}
\mathcal{G}(R,t).
\label{eq32}
\end{equation}
Assuming each burst lasts for $t_{\mathrm{burst}}\simeq t_{\mathrm{dyn}}$, the time-averaged sBH heating flux is
\begin{equation}
Q_{\mathrm{sBH}}^{0}(R,t)
=
\dot{\mathcal{N}}_{\mathrm{rec}}
Q_{\mathrm{event}}^{\mathrm{peak}}
t_{\mathrm{burst}} .
\label{eq33}
\end{equation}
The heating ratio
\begin{equation}
\Phi(R,t)
=
\frac{
Q_{\mathrm{sBH}}^{0}(R,t)
}{
Q_{\mathrm{vis}}^{0}(R)
}
\label{eq34}
\end{equation}
is used below to quantify where sBH-driven reconnection becomes dynamically important.

The instantaneous heating is realized as a shot-noise process around this mean. At each time step, we draw $N_{\rm event}\sim{\rm Poisson}[\Lambda(R,t)]$, where $\Lambda(R,t)=\dot {\mathcal{N}}_{\rm rec}(R,t)\delta t$. A stochastic burst prescription is appropriate because magnetic reconnection in turbulent plasma is intrinsically intermittent and produces burst-like energy release \cite[e.g.,][]{2019PhRvL.122e5101Z,2023ApJ...946L...3T}. The heating evolves as
\begin{multline}
Q_{\mathrm{sBH}}^{+}(R,t_{n+1})
=
Q_{\mathrm{sBH}}^{+}(R,t_{n})
\exp
\left(
-\frac{\delta t}{t_{\mathrm{dyn}}}
\right)
\\
+
\sum_{k=1}^{N_{\mathrm{event}}}
A_{k}
Q_{\mathrm{event}}^{\mathrm{peak}}(R,t_{n}),
\label{eq35}
\end{multline}
where the event amplitudes follow a unit-mean lognormal distribution,
\begin{equation}
A_{k}
=
\exp
\left(
\sigma_{\mathrm{amp}}z
-
\frac{1}{2}\sigma_{\mathrm{amp}}^{2}
\right),
\label{eq36}
\end{equation}
where $z\sim\mathcal{N}(0,1)$ is the Gaussian random number. This form captures the intermittency of reconnection-driven energy release \cite[e.g.,][]{2019PhRvL.122e5101Z,2023ApJ...946L...3T}. When $\Lambda\ll1$, isolated flares dominate the heating; when $\Lambda\gg1$, many events overlap and form a smoother heating floor. We note that the population-balance equation, Equation~(\ref{eq5}), describes the statistical (ensemble-averaged) distribution of embedded sBHs, so that $N_R(R,t)$ represents the expected number of objects per annulus rather than a specific integer realization. Near a migration trap, the local population can be modest, and the discreteness of individual reconnection sites is therefore physically relevant. This discreteness is not lost in our treatment: it is captured at the light-curve level by the shot-noise prescription of Equations~(\ref{eq35})-(\ref{eq36}), in which the number of events per time step is drawn from a Poisson distribution with mean $\Lambda=\dot{\mathcal{N}}_{\rm rec}\delta t$ and the amplitudes from a lognormal distribution. In the small-$\dot{\mathcal{N}}_{\rm rec}$ (low-$\Lambda$) regime, this naturally produces isolated, well-separated flares rather than a smooth heating term, so that the burst-like character of a sparsely populated trap is preserved. The population-balance equation thus supplies the mean event rate, while the stochastic realization restores the intermittency expected from a finite, discrete population.

\subsubsection{Thermal Response}
\label{subsubsec:thermal_response}

Gap opening affects the disk thermal balance not only through the reduced co-moving density that enters the reconnection trigger Equation~(\ref{eq26}), but also through the background disk structure itself. In the sBH-active disk, the same gap-attenuation factor $f^{\rm 1D}_{\rm gap}(R,t)$ scales the local background heating rates, so that the total heating driving the local thermal evolution is
\begin{equation}
Q_{\mathrm{heat}}^{+}=
f^{\rm 1D}_{\rm gap}\left(Q^{+}_{\rm vis}+Q^{+}_{\rm mc}\right)
+f_{\rm int}\,Q^{+}_{\rm sBH},
\label{eq37}
\end{equation}
where $f_{\rm int}=0.9$ is the fraction of reconnection power deposited into the disk interior and the remaining fraction is treated as surface-layer heating (see Equation~(\ref{eq40})). Because the embedded population is initially absent, $f^{\rm 1D}_{\rm gap}\to1$ at early times and Equation~(\ref{eq37}) reduces to the CHAR-like background; gaps develop only as the trapped population grows. The mid-plane temperature evolves as
\begin{equation}
\Pi C
\frac{d\ln T_{\mathrm{c}}}{dt}
=
Q_{\mathrm{heat}}^{+}
-
Q_{\mathrm{rad}}^{-}.
\label{eq38}
\end{equation}

Gap opening enters this balance through a single depleted surface density,
\begin{equation}
\Sigma_{\rm B}(R,t)=f^{\rm 1D}_{\rm gap}(R,t)\,\Sigma(R),
\label{eq39}
\end{equation}
which is propagated self-consistently through all background thermal quantities: the vertically integrated pressure $\Pi$, the scale height $H$, the optical depth $\tau_\mathrm{d}$, the radiative cooling $Q^{-}_{\rm rad}$, the vertically integrated gas pressure $\Pi_{\rm gas}$, the pressure ratio $\beta_{\rm gas}=\Pi_{\rm gas}/\Pi$, and the dimensionless heat capacity $C=C_{0}[\beta_{\rm gas}(R,t)]$ are all evaluated at $\Sigma_{\rm B}$ rather than at the unperturbed $\Sigma$, with $C_{0}(\beta_{\rm gas})$ the standard-disk heat capacity at the same $\beta_{\rm gas}$ (see Equation~(10) in \cite{2020ApJ...891..178S}). $\Sigma_{\rm local}$ uses the un-floored gap depletion for accretion, whereas $\Sigma_{\rm B}$ uses the floored one-dimensional depletion $f_{\rm gap}^{1D}$ for the thermal calculation. Gap opening therefore has two competing consequences for the variability: by lowering $\rho_{c}$ it suppresses the reconnection power ($P_{m}\propto P_{r}\propto\rho_{c}$), while by lowering the gas-pressure contribution to $\Pi$ it reduces the local thermal inertia $\Pi C$. For a fixed absolute sBH perturbation $f_{\rm int}Q^{+}_{\rm sBH}$, the fractional temperature response $\propto 1/(\Pi C)$ is enhanced in depleted regions, so a gap-bearing trap responds more sharply to a given reconnection event even as its background heating is reduced. Gap opening is thus a self-regulating element of the thermal response, not a simple multiplicative correction to the heat capacity alone.

After solving Equation~\eqref{eq38}, the emergent surface flux is
\begin{equation}
Q_{\mathrm{surf}}
=
Q_{\mathrm{rad}}^{-}
+
\left(1-f_{\mathrm{int}}\right)
Q_{\mathrm{sBH}}^{+},
\label{eq40}
\end{equation}
and the effective temperature is
\begin{equation}
T_{\mathrm{eff}}
=
\left(
\frac{
Q_{\mathrm{surf}}
}{
2\sigma_{\mathrm{SB}}
}
\right)^{1/4}.
\label{eq41}
\end{equation}
The resulting $T_{\mathrm{eff}}(R,t)$ is used to synthesize the optical/UV light curves in Section~\ref{subsec:lightcurve}.

The reconnection-heating model involves several microphysical parameters that are not tightly fixed by first principles: the effective ram-to-magnetic pressure ratio $\beta_{p}=0.01$ \cite[e.g.,][]{2018ApJ...852...95N,2022MNRAS.513.4267N}; the plasma magnetization $\sigma=15$ (with a plausible range of $3$-$25$) \cite[e.g.,][]{2013MNRAS.431..355G,2021JPlPh..87e9012R,2022ApJ...924L..32R}; the current-sheet aspect ratio $g_{s}=0.12$ \cite[e.g.,][]{2021JPlPh..87e9012R,2022ApJ...924L..32R,2025ApJ...991..167X}; the interior-deposition fraction $f_{\rm int}=0.9$; and the lognormal burst amplitude $\sigma_{\rm amp}=0.8$. We adopt values that are either standard in the relativistic-reconnection literature or deliberately conservative. 

\begin{deluxetable}{llc} \tablecaption{Model parameters that may affect the predicted variability signatures.} 
\tablehead{ \colhead{Parameter name} & \colhead{Symbol} & \colhead{Fiducial value / explored values} } \startdata \multicolumn{3}{c}{Disk and CHAR background} \\
 \hline SMBH mass & $M_\bullet$ & $5\times10^7\,M_\odot$ \\
 Viscosity & $\alpha$ & 0.2 \\
 Eddington ratio & $\dot m$ & 0.02, 0.05, 0.20 \\
 Coronal fraction & $k_{\rm ratio}$ & $1/3$ \\
 \hline \multicolumn{3}{c}{Migration and gap-opening prescription} \\
 \hline Type-I factor & $C_{\rm I}$ & 3.8 \\
 Gap parameter & $K'$ & $0.04q^2/(h^5\alpha)$ \\
 Accretion trap & $K'_{\rm acc}$ & $0.01(h/0.05)$ \\
 Accretion lower edge & $K'_{\rm acc,l}$ & $K'_{\rm acc}10^{-0.5}$ \\
 Gap onset & $K'_{\rm gap}$ & 50 \\
 Deep-gap onset & $K'_{\rm gap,h}$ & 250 \\
 Gap floor & $f_{\rm gap,min}$ & 0.5 \\
 Speed cap & $|v_{\rm R}|_{\rm max}$ & $0.1R\Omega_K$ \\
 \hline \multicolumn{3}{c}{Embedded sBH population and source terms} \\
 \hline Initial embedded number & $N_{\rm ref}$ & 0 \\
 Capture slope & $\gamma_{\rm cap}$ & 1.5 \\
 Inner cutoff & $R_{\rm in,cut}$ & $10R_{\rm S}$ \\
 In-situ width & $\sigma_{\rm insitu}$ & 0.15 \\
 Capture MF slope & $\alpha_{\mathrm{cap}}$ & 1.0 \\
 In-situ MF slope & $\alpha_{\mathrm{M}}$ & 1.3 \\
 Mass range & $M_{\rm min}, M_{\rm max}$ & $8M_\odot,\ 1000M_\odot$ \\
 BH yield & $\eta_{\rm BH}$ & $0.03\,M_\odot^{-1}$ \\
 Star formation rate & $\dot M_{\rm sf}$ & $0.05\,M_\odot\,{\rm yr^{-1}}$ \\
 Capture rate & $\dot N_{\rm cap}$ & $10^3\,{\rm Myr^{-1}}$ \\
 Import time & $t_{\rm imp}$ & $10^0,\ 10^2,\ 5\times10^4,\ 5\times10^5\,{\rm yr}$ \\
 \hline \multicolumn{3}{c}{Magnetic reconnection and stochastic heating} \\
 \hline Magnetization & $\sigma$ & 15 $(3$--25) \\ 
 Pressure ratio & $\beta_{\rm p}$ & 0.01 \\ 
 Sheet width & $g_s$ & 0.12 \\
 Interior fraction & $f_{\rm int}$ & 0.9 \\
 Amplitude scatter & $\sigma_{\rm amp}$ & 0.8 \\
 Burst time & $t_{\rm burst}$ & $\simeq t_{\rm dyn}=H/v_A$ \\
 \hline \multicolumn{3}{c}{Synthetic light curves and diagnostics} \\
 \hline Simulation length & $t_{\rm sim}$ & $2\times10^4\,{\rm days}$ \\
 Cadence & $\delta t$ & 0.5 day \\
 Burn-in & $t_{\rm burn}$ & 500 days \\
 Bands & $\lambda$ & 1367, 1928, 2600, 3000, 5100 ${\rm \AA}$ \\
 Reference band & $\lambda_0$ & 1367 ${\rm \AA}$ \\
 Coherence cut & $C_{\rm coh}$ & $>0.5$ \\
 \enddata \tablecomments{Only parameters that are adopted, scanned, or able to affect the predicted population evolution and variability diagnostics are listed. Quantities such as $\gamma$, $\tau_{\rm DRW}$, $\beta_{\rm lag}$, and $C_{\rm coh}$ are diagnostic outputs rather than physical input parameters, except where their measurement thresholds or fitting ranges are specified above. } 
 \label{tab:parameters} 
\end{deluxetable}

\subsection{Synthetic Optical/UV Light Curves and Diagnostics}
\label{subsec:lightcurve}

With the time-dependent effective temperature $T_{\mathrm{eff}}(R,t)$ obtained from the thermal response calculation, we synthesize optical/UV light curves by integrating the local blackbody emission over the disk surface,
\begin{equation}
\lambda L_{\lambda}(t)
=
4\pi^{2}\lambda
\int_{R_{\mathrm{in}}}^{R_{\mathrm{out}}}
R\,dR\,
B_{\lambda}
\left[
T_{\mathrm{eff}}(R,t)
\right],
\label{eq42}
\end{equation}
where $B_{\lambda}(T)$ is the Planck function. The factor $4\pi^{2}$ accounts for two-sided disk emission integrated over azimuth. The radial integral is evaluated numerically on the same logarithmic grid used for the disk and population calculations.

We compute light curves in five representative rest-frame bands spanning the UV to optical range: $1367~\mathrm{\AA}$, $1928~\mathrm{\AA}$, $2600~\mathrm{\AA}$, $3000~\mathrm{\AA}$, and $5100~\mathrm{\AA}$. The UV bands mainly trace the hotter inner disk, where sBH-driven reconnection heating is expected to produce the strongest and most rapid response. The optical bands arise from cooler and typically larger disk radii, and therefore encode both the radial redistribution of thermal perturbations and the smoothing effect of the extended emitting area. This wavelength coverage allows us to track how localized sBH heating propagates into observable optical/UV variability.

The reference state is defined at the imported time (i.e., $t_{\mathrm{imp}}=10^{0}~\mathrm{yr}$) and corresponds to the CHAR-like background disk without sBH-driven reconnection heating, i.e., $Q_{\mathrm{sBH}}^{+}(R,t)=0$. We then compare this baseline with models that include the evolved sBH populations imported at later times. The difference between the CHAR-like reference and the evolved cases isolates the optical/UV signatures of embedded sBHs, including enhanced short-time-separation variability, changes in the structure-function (SF) shape, and modifications of the inter-band time-delay behavior.

We characterize the optical/UV variability using the SF, computed with a segment-averaged normalized median absolute deviation (NMAD) estimator, following the treatment adopted for simulated CHAR light curves \citep[e.g.,][]{2020ApJ...891..178S}. Each light curve is first converted to magnitudes, $m(t)=-2.5\log_{10}\lambda L_{\lambda}(t)$, and divided into $N_{\mathrm{seg}}=5$ contiguous, equal-length segments. Within each segment, and for a set of logarithmically spaced time separations $\Delta t$, we collect all magnitude differences $\Delta m_{ij}=m(t_{j})-m(t_{i})$ with $\Delta t=|t_{j}-t_{i}|$, and estimate the SF from their NMAD,
\begin{equation}
{\rm SF}(\Delta t)
=
1.48 \times
{\rm median}\!\left(
\left|
\Delta m_{ij}
-
{\rm median}(\Delta m_{ij})
\right|
\right),
\label{eq43}
\end{equation}
where the factor $1.48$ normalizes the NMAD to the standard deviation for Gaussian-distributed differences. The final SF at each $\Delta t$ is obtained by averaging the segment SFs. This NMAD estimator is less sensitive to outliers than a simple rms estimator and is therefore better matched to the SF treatment used in CHAR light-curve analyses.

We quantify the short-to-intermediate time-separation SF shape by fitting a single power-law form,
\begin{equation}
{\rm SF}(\Delta t) \propto \Delta t^{\gamma},
\label{eq44}
\end{equation}
to the 3000\,\AA\ SF. The slope $\gamma$ is obtained by ordinary least-squares regression of $\log_{10}\mathrm{SF}$ on $\log_{10}\Delta t$ over $\Delta t = 14$--$100$ days. This time-separation range is chosen to match the window over which the low-luminosity AGN SF slopes of \cite{2026MNRAS.tmp..992T} were fitted, enabling a like-for-like comparison in Section~\ref{subsubsec:state_properties}. Because the SF slope can depend on the fitted $\Delta t$ interval, the measured $\gamma$ should be regarded as an effective slope over 14--100 days rather than as a global property of the full light curve.

The DRW damping timescale is obtained by fitting the DRW structure-function model
\begin{equation}
{\rm SF}_{\mathrm{DRW}}(\Delta t)
=
{\rm SF}_{\infty}
\left[
1-\exp\left(
-\frac{\Delta t}{\tau_{\mathrm{DRW}}}
\right)
\right]^{1/2},
\label{eq45}
\end{equation}
to the same NMAD SF, where ${\rm SF}_{\infty}$ is the asymptotic SF amplitude and $\tau_{\mathrm{DRW}}$ is the effective damping timescale. For this fit, the segment-averaged SF is measured for time separations up to one segment length: dividing each $t_{\rm sim}=2\times10^{4}$ day light curve (after removing the first $t_{\rm burn}=500$ days) into $N_{\rm seg}=5$ equal segments gives a maximum accessible separation of $\approx4\times10^{3}$ days. The DRW SF model is then fitted over $\Delta t = 100$ days up to this per-segment maximum. Because the damping turnover occurs at $\tau_{\rm DRW}\sim10^{2}$--$10^{3}$ days, well within this baseline, the fitted timescale is robustly constrained. The two parameters $(\tau_{\mathrm{DRW}},{\rm SF}_{\infty})$ are optimized in logarithmic space using a robust soft-$\ell_{1}$ loss applied to the residuals between $\log{\rm SF}_{\mathrm{DRW}}$ and $\log{\rm SF}$, which down-weights discrepant SF points relative to a least-squares fit. In this analysis, $\tau_{\mathrm{DRW}}$ is derived from this SF fit rather than from an autocorrelation-function exponential fit, and should be interpreted as an effective DRW-SF timescale over the fitted interval. A fit is flagged as unreliable when the best-fit $\tau_{\mathrm{DRW}}$ lies too close to the upper edge of the fitted range ($\tau_{\mathrm{DRW}}>0.9\,\Delta t_{\mathrm{max}}$) or below $\Delta t_{\mathrm{min}}/3$, since in those cases the SF has not clearly turned over within the fitting window. As for the SF slope, the fitted range is stated explicitly in the corresponding figure title.

For inter-band time-delay measurements, we use the frequency-domain phase-lag spectrum. For a pair of light curves, we estimate the cross-spectrum $S_{xy}(f)$ using a Welch-style averaged periodogram with Hann-windowed, $50\%$-overlapping segments, where the segment length is chosen to span several cycles at the lowest analyzed frequency. Prior to the transform, each light curve is mean-subtracted and normalized by its standard deviation. The cross-spectral phase $\phi(f)=\arg S_{xy}(f)$ is converted into a frequency-dependent time delay through
\begin{equation}
\tau(f)=-\frac{\phi(f)}{2\pi f},
\label{eq46}
\end{equation}
where the sign convention is chosen such that a positive lag means that the longer-wavelength band lags behind the shorter-wavelength reference band. At each frequency the squared coherence $C_{\rm coh}(f)=|S_{xy}(f)|^{2}/[S_{xx}(f)\,S_{yy}(f)]$ quantifies the degree of linear correlation between the two bands. The band-averaged lag is computed as a weighted mean of $\tau(f)$ over the chosen frequency interval, with weights given by the product of the coherence and the cross-spectral amplitude, $w(f)=C_{\rm coh}(f)\,|S_{xy}(f)|$, so that frequencies that are both coherent and energetic dominate the estimate.

For the $5100~\mathrm{\AA}$ versus $3000~\mathrm{\AA}$ delay we average $\tau(f)$ over $1/70<f<1/3~\mathrm{day}^{-1}$ and require a mean coherence above $0.5$. The wavelength-dependent lag analysis uses the $1367~\mathrm{\AA}$ band as the reference and measures lags for the longer-wavelength bands over the lower-frequency interval $1/300<f<1/30~\mathrm{day}^{-1}$, with a less restrictive mean-coherence threshold of $0.4$, since at these lower frequencies the disk thermal response is expected to be more relevant. The two diagnostics deliberately probe different frequency ranges: the $5100/3000$~\AA{} delay is averaged over $1/70<f<1/3~{\rm day}^{-1}$ to match the day-to-week timescales sampled by continuum reverberation campaigns, whereas the wavelength-dependent exponent uses the lower-frequency interval $1/300<f<1/30~{\rm day}^{-1}$, where the slower, large-scale thermal response of the disk dominates. Because the phase lag of a reprocessing disk is in general frequency-dependent, the two measurements are complementary and are not required to lie on a single $\tau(\lambda)$ curve. A lag measurement is marked unreliable if the mean coherence falls below the adopted threshold, if the inferred lag is consistent with zero, or if the frequency-dependent lag scatter is too large compared with the mean lag (specifically, if its standard deviation exceeds a few times the mean lag).

The wavelength dependence of the disk continuum lags is then summarized by fitting a power law,
\begin{equation}
\tau(\lambda)
=
\tau_{0}
\left[
\left(
\frac{\lambda}{\lambda_{0}}
\right)^{\beta_{\mathrm{lag}}}
-
1
\right],
\label{eq47}
\end{equation}
where $\lambda_{0}=1367~\mathrm{\AA}$ is the reference wavelength, $\tau_{0}$ is the lag normalization relative to this reference band, and $\beta_{\mathrm{lag}}=4/3$ corresponds to the standard thin-disk expectation $\tau\propto\lambda^{4/3}$. In practice, we measure the lag of each longer-wavelength band relative to $1367~\mathrm{\AA}$ with the phase-lag method described above, retain only bands with positive, sufficiently coherent lags, and obtain $\beta_{\mathrm{lag}}$ from a linear regression of $\log\tau$ on $\log(\lambda/\lambda_{0})$. Departures from $\beta_{\mathrm{lag}}=4/3$ indicate that the radial temperature response has been modified by localized sBH heating, gap-induced thermal-inertia changes, or both.

All calculations are implemented in MATLAB. The sBH population equation is solved over $5\times10^{5}~\mathrm{yr}$ using an operator-split finite-volume scheme with absorbing radial boundaries \cite[e.g.,][]{1999MNRAS.307...79I}. Population snapshots at selected import times are then passed to the magnetic-heating module to compute $Q_{\mathrm{sBH}}^{0}(R,t)$ and the corresponding shot-noise realization $Q_{\mathrm{sBH}}^{+}(R,t)$. The optical/UV light-curve simulations typically span $t_{\mathrm{sim}}=2\times10^{4}~\mathrm{days}$ with a cadence of $\delta t=0.5~\mathrm{day}$, and the first $t_{\mathrm{burn}}=500~\mathrm{days}$ are discarded before measuring the variability diagnostics.

\section{Results}\label{sec:results}
\subsection{Migration-trap Formation}\label{subsec:migration}

\begin{figure*}[htbp]
    \centering
    \includegraphics[width=0.3\textwidth]{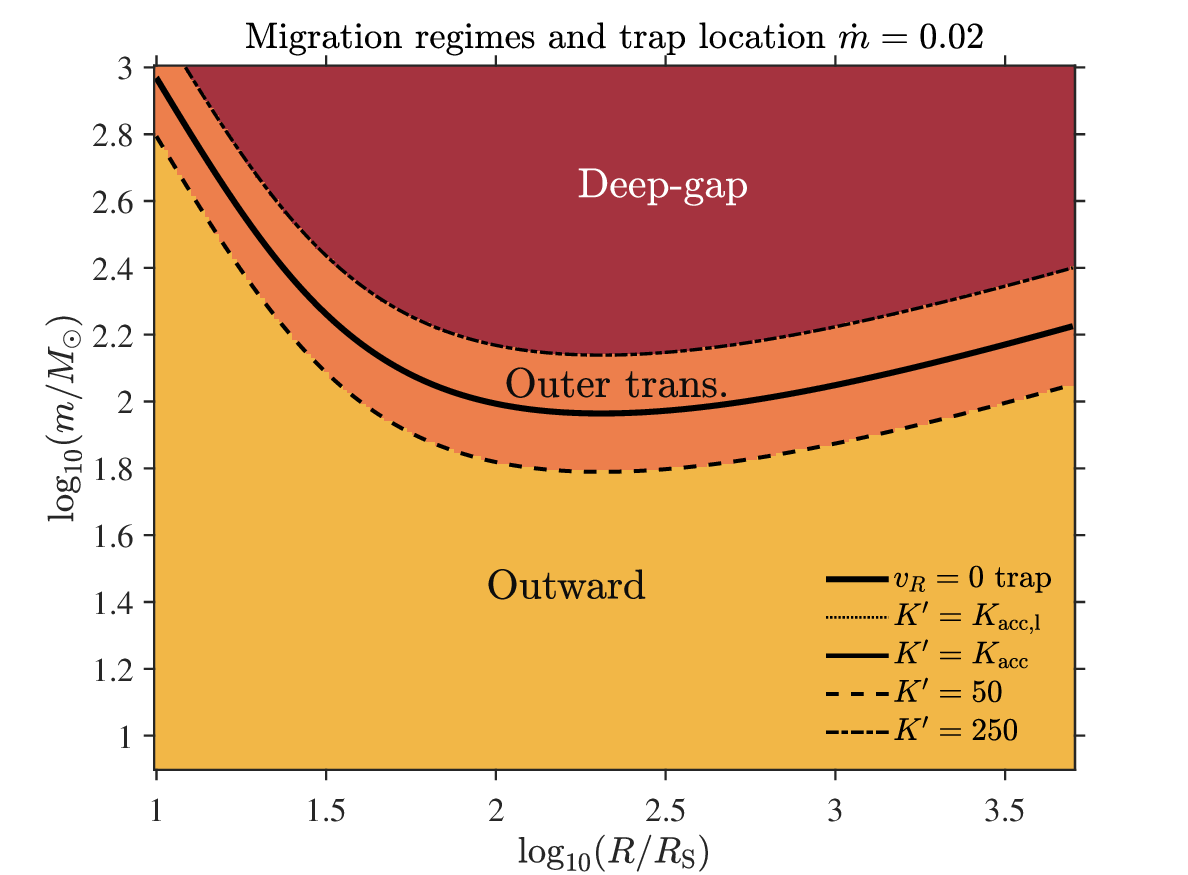}
    \includegraphics[width=0.3\textwidth]{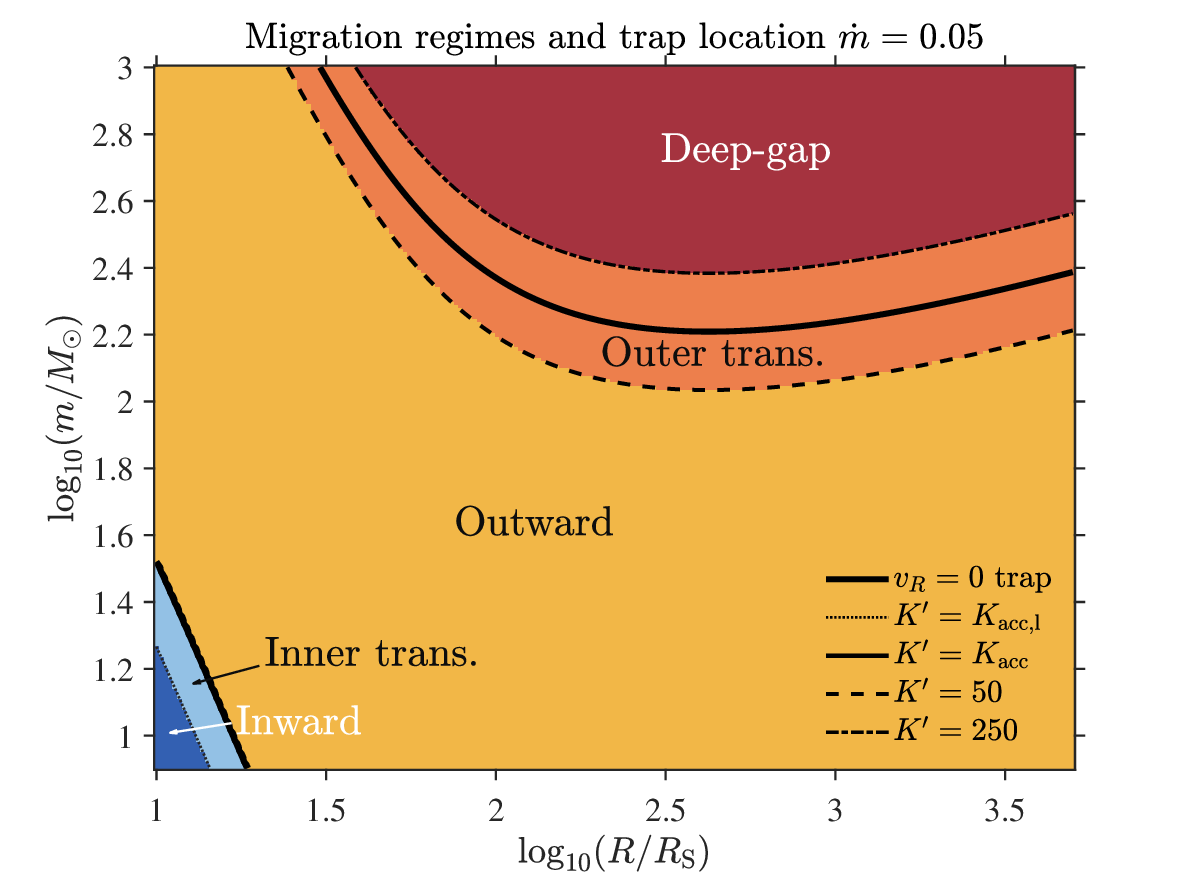}
    \includegraphics[width=0.3\textwidth]{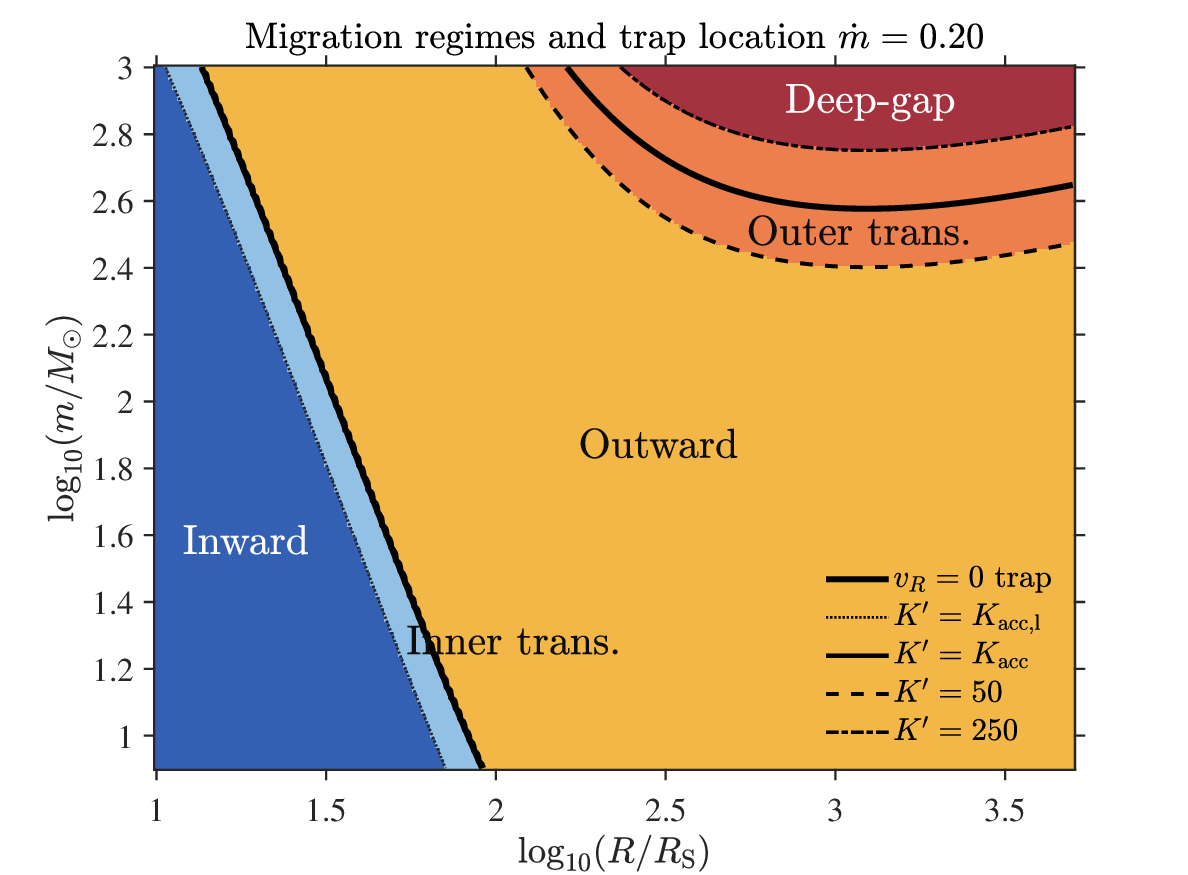}
    \caption{Migration regime maps in the $M$-$R$ state space for different Eddington ratios. Left panel: $\dot{m}=0.02$; middle panel: $\dot{m}=0.05$; right panel: $\dot{m}=0.20$. The color regions represent the dominant migration states: blue/light blue indicates inward migration (including Type I migration and transition zone), yellow/orange indicates accretion-driven outward migration (and transition zone), and deep red indicates inward migration of deep gaps. The black solid line ($v_\mathrm{R} = 0$) indicates the outward migration boundary with zero net velocity (i.e., the migration trap). The results show that as the accretion rate increases, the disk aspect ratio increases, resulting in a significant expansion of the outward migration zone, and subsequently pushing the migration trap to a larger radius and higher sBH mass region.}
    \label{fig1}
\end{figure*}

We first examine how the embedded sBH population evolves under different global accretion states of the AGN disk. The population equation is evolved for $5\times10^{5}~\mathrm{yr}$ with absorbing inner and outer radial boundaries. We compare three Eddington ratios, $\dot{m}=0.02$, $0.05$, and $0.20$, in order to isolate how the background disk structure controls migration and trapping.

The migration behavior is primarily regulated by the gap-opening parameter $K'$. Because $K'$ depends sensitively on the disk aspect ratio, $K'\propto h^{-5}$, changes in the accretion rate reshape the migration map in the $(R,M)$ phase space. Figure~\ref{fig1} shows the corresponding migration regimes for the three accretion states. At low and moderate accretion rates, the disk is relatively thin, so embedded sBHs more easily open gaps and enter the inward gap-modified migration regime. The outward-migration zone is therefore radially narrow. In contrast, when $\dot{m}=0.20$, the inner disk becomes geometrically thicker because of enhanced radiation pressure, suppressing gap opening and expanding the region where accretion-driven positive torques can operate \cite[e.g.,][]{1973A&A....24..337S,1988ApJ...332..646A,2019ApJ...885..144J}. The null-velocity curves, where $v_{\mathrm{R}}=0$, define the migration traps. These curves act as phase-space attractors because sBHs migrating from opposite sides are driven toward the same zero-velocity boundary.

\begin{figure*}[htbp]
    \centering
    \includegraphics[width=0.4\textwidth]{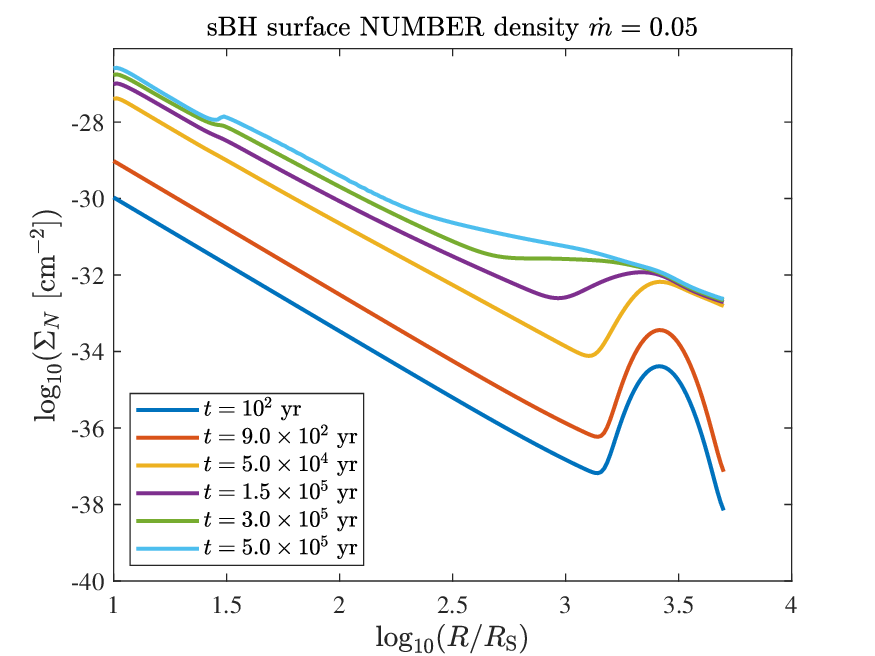}
    \includegraphics[width=0.4\textwidth]{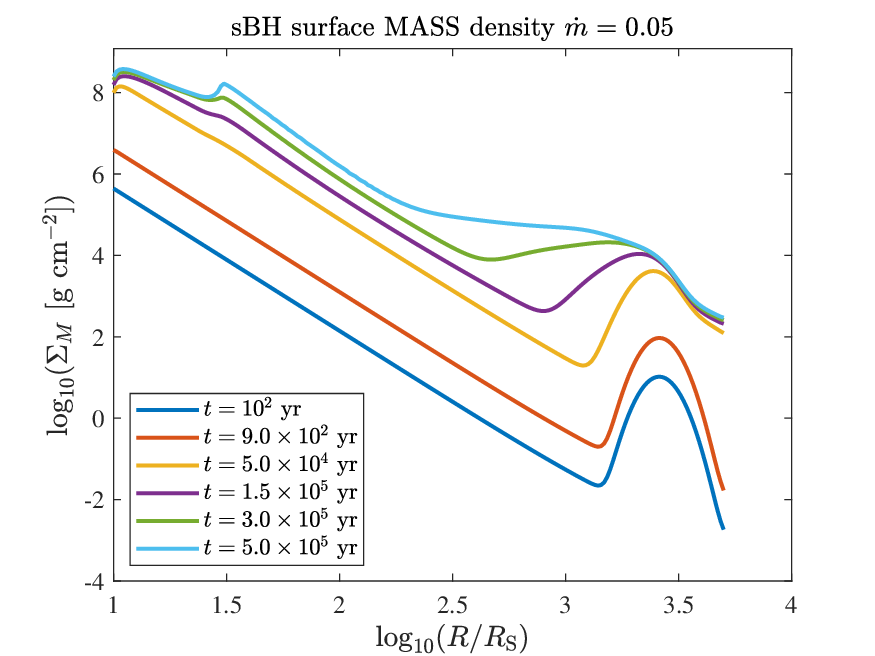}
    \includegraphics[width=0.4\textwidth]{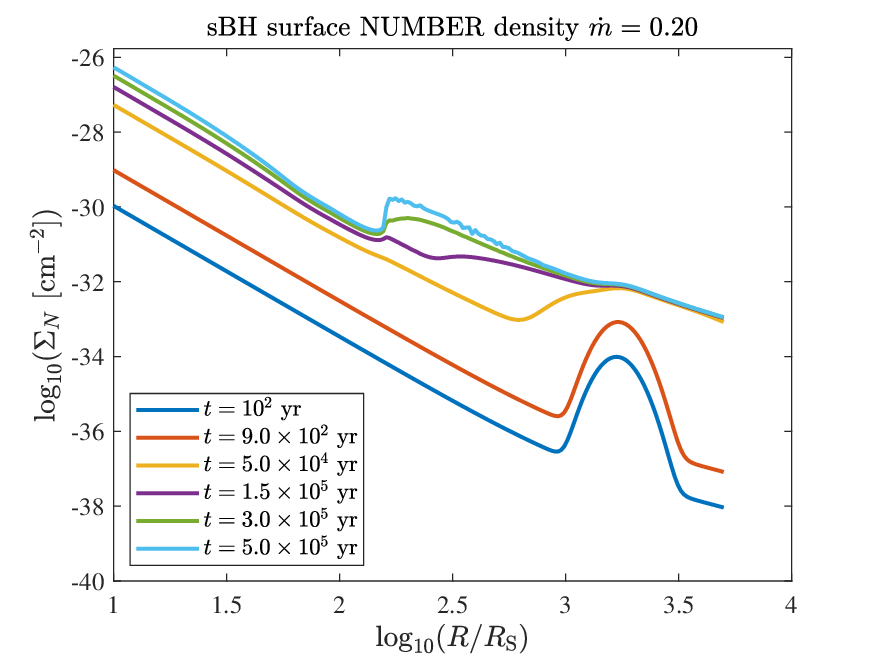}
    \includegraphics[width=0.4\textwidth]{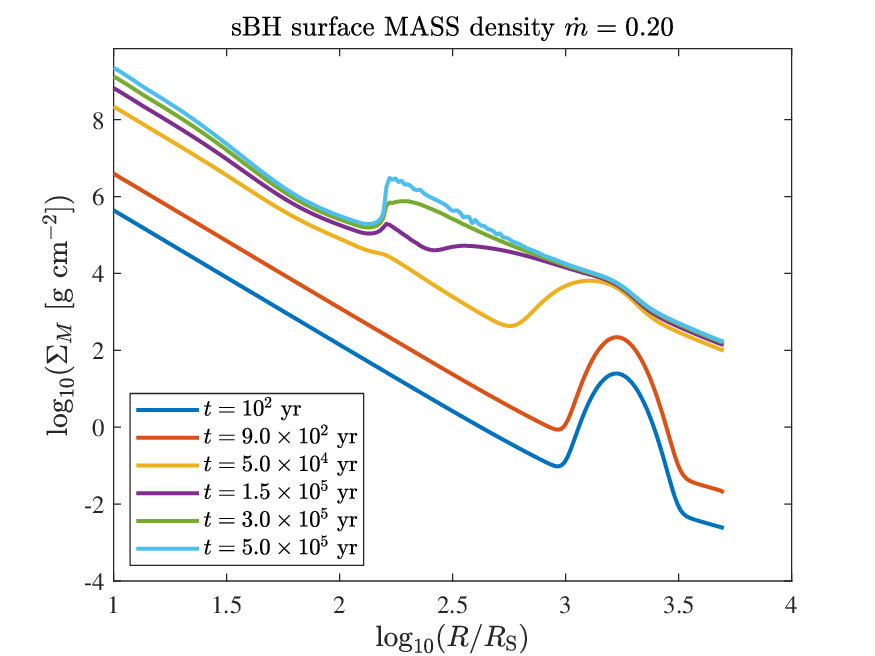}
    \caption{Radial evolution of the sBH surface number density and surface mass density over $5\times10^{5}~\mathrm{yr}$. Top panels: $\dot{m}=0.05$; bottom panels: $\dot{m}=0.20$. In the moderate-accretion model, a secondary peak forms near $\log_{10}(R/R_{\mathrm{S}})\simeq1.5$, corresponding to the migration trap identified in Figure~\ref{fig1}. In the high-accretion model, no comparably sharp secondary trap peak appears. The larger disk aspect ratio weakens gap opening and dilutes the embedded-sBH perturbation in the vertically/azimuthally averaged disk density, so the radial distribution is dominated by inner-disk accumulation rather than by a distinct trap-induced enhancement. This difference sets the accretion-state dependence of the subsequent sBH-driven magnetic heating.}
    \label{fig2}
\end{figure*}

Figure~\ref{fig2} shows the resulting radial pile-up of the embedded population. For $\dot{m}=0.05$, the profiles of both the surface number density $\Sigma_{\mathrm{N}}$ and surface mass density $\Sigma_{\mathrm{M}}$ show a pronounced secondary enhancement near $\log_{10}(R/R_{\mathrm{S}})\simeq1.5$. This radius coincides with the null-velocity boundary in the migration map, confirming that the feature is a genuine trap-induced pile-up rather than a remnant of the initial condition. For $\dot{m}=0.20$, the radial profiles are dominated by an inner-disk enhancement and no comparably sharp secondary trap peak appears. This does not necessarily mean that the migration trap is absent. Rather, the larger disk aspect ratio weakens gap opening and reduces the surface-density contrast produced by embedded sBHs in the vertically and azimuthally averaged disk. As a result, the trap-induced feature is smeared out in the high-accretion model.

The surface mass density enhancement is especially important for the subsequent variability calculation. Because magnetic reconnection heating depends on both the number of available sBHs and their characteristic mass, a trap-induced increase in $\Sigma_{\mathrm{N}}$ and $\Sigma_{\mathrm{M}}$ produces a localized reservoir of potential heating events. However, the same accumulation also strengthens gap opening and can reduce the local gas density available for reconnection. Thus, the migration trap does not simply act as a steady heat source; it creates a localized, self-regulated region where sBH abundance, gas depletion, and magnetic heating compete. This spatially nonuniform population provides the initial condition for the sBH-driven heating model discussed in the next subsection.

\subsection{Radial Structure of sBH-driven Heating}
\label{subsec:radial_heating}

We next examine how the migration-driven sBH distribution is mapped into disk heating. This step provides the bridge between the population synthesis and the subsequent optical/UV light-curve calculation. For each accretion state, the sBH population is evolved for $5\times10^{5}~\mathrm{yr}$ with absorbing radial boundaries. Objects that cross the inner boundary are removed from the disk, while the outer boundary is treated as outflow-only. Continuous replenishment by NSC capture and in-situ formation is included through the source terms described in Section~\ref{subsubsec:initial_conditions}. The resulting population snapshots are then imported into the magnetic-heating calculation at four representative evolutionary times, $t_{\mathrm{imp}}=10^0,~10^{2},\ 5\times10^{4},\ \mathrm{and}\ 5\times10^{5}~\mathrm{yr}$. The $t_{\mathrm{imp}}=10^0$ yr case is used as a CHAR-like no-sBH reference, while the remaining three epochs correspond respectively to an early weak-pile-up phase, an intermediate accumulation phase, and a late trap-regulated phase.

At each imported epoch, the time-averaged sBH heating profile is computed from the local values of $N_{\mathrm{R}}(R,t)$, $\langle M_{\mathrm{sBH}}\rangle(R,t)$, and the gap attenuation factor $f_{\mathrm{gap}}^{\mathrm{1D}}(R,t)$. We use the heating ratio $\Phi(R,t)$, defined in Equation~\eqref{eq34}, to quantify the local importance of sBH-driven reconnection relative to viscous heating. $\Phi\ll1$ indicates a disk thermally dominated by standard viscous dissipation, while $\Phi\sim1$ indicates that sBH-driven reconnection can locally compete with the intrinsic disk heating.

\begin{figure}
    \centering
    \includegraphics[width=0.48\textwidth]{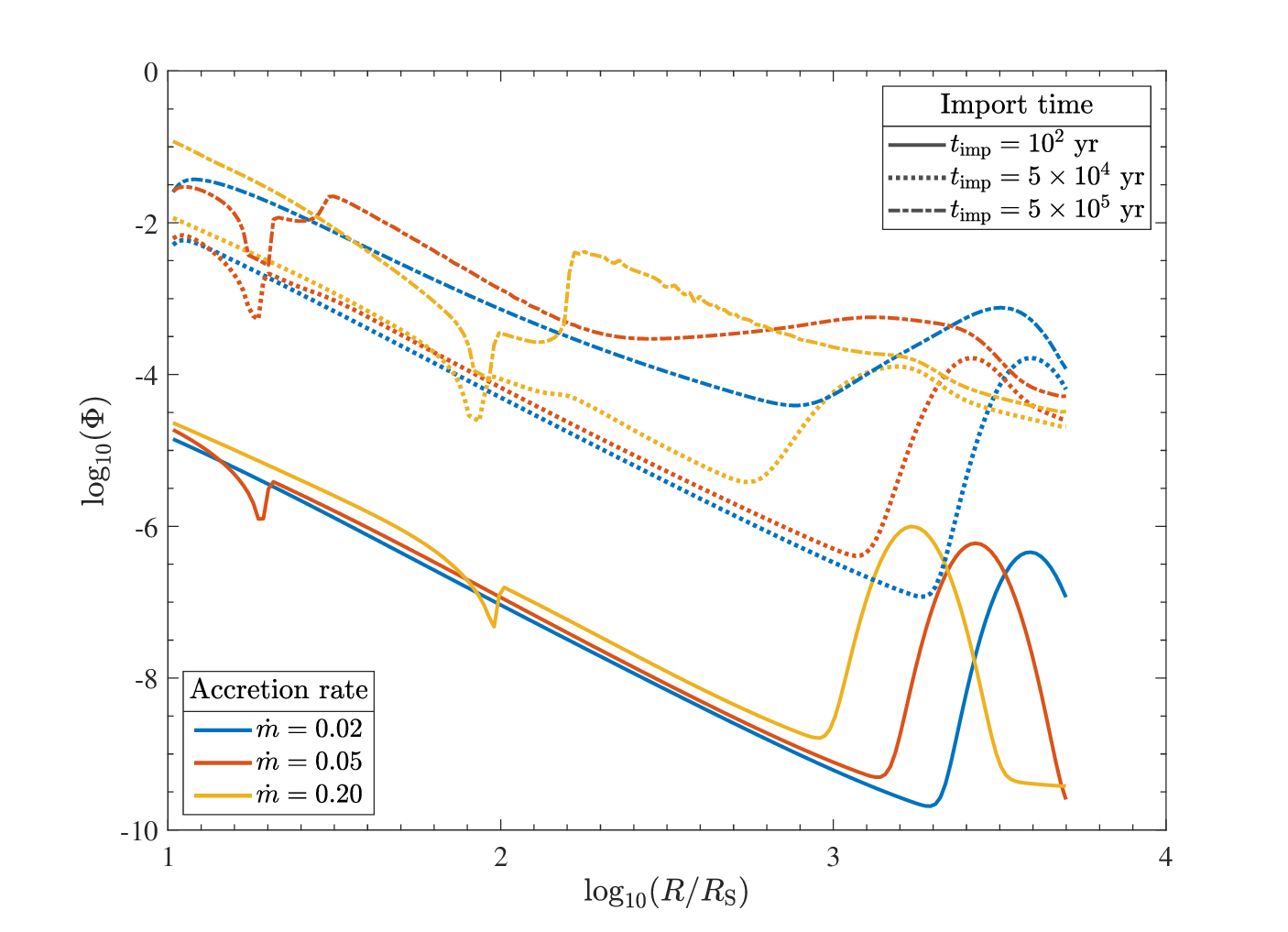}
    \caption{Radial profile of the sBH heating ratio $\Phi(R,t)=Q_{\mathrm{sBH}}^{0}(R,t)/Q_{\mathrm{vis}}^{0}(R)$ for different accretion rates and sBH population ages. Colors denote different accretion rates, $\dot{m}=0.02$, $0.05$, and $0.20$, while line styles denote imported population times $t_{\mathrm{imp}}=10^{2}$, $5\times10^{4}$, and $5\times10^{5}~\mathrm{yr}$. At early times, the sBH population had not yet accumulated efficiently and $\Phi\ll1$ over most radii. As migration proceeds, sBH pile-ups enhance the reconnection heating in the inner disk, especially for the low and moderate accretion states. At late times, local depressions in $\Phi$ can appear near the trap because gap opening reduces the gas density available for ram-pressure-driven reconnection. The high-accretion model shows weaker fractional heating because the stronger viscous background dilutes the localized sBH contribution.}
    \label{fig3}
\end{figure}

Figure~\ref{fig3} shows $\Phi(R,t)$ for the three accretion states and the three sBH-active imported population ages. At the earliest epoch, $\Phi$ remains well below unity over most radii. This reflects the fact that the initial sBH population has not yet been dynamically concentrated by migration traps, so both the event rate and the characteristic reconnection power are small. The disk is therefore close to the CHAR-like background state, with only weak additional heating from embedded sBHs.

As the population evolves, the radial heating profile becomes strongly structured. For all accretion states, sBH migration produces a substantial increase in $\Phi$ over the inner disk. In these models, the sBH number density grows near the trap and the local event rate increases accordingly. Outside the most depleted part of the gap, the gas density remains high enough for ram-pressure-triggered reconnection to operate efficiently. These radii are therefore expected to be the most important sites for sBH-induced optical/UV variability.

The heating profile, however, does not simply trace the sBH number density. A prominent feature of Figure~\ref{fig3} is that the late-time profiles can develop a local depression near the trap. This behavior is a direct consequence of gap feedback. As sBHs accumulate and grow, they open deeper gaps and reduce the co-moving gas density through $f_{\mathrm{gap}}^{\mathrm{1D}}$. Since the ram pressure scales as $P_{\mathrm{r}}\propto\rho_{\mathrm{c}}v_{\mathrm{rel}}^{2}$, a lower $\rho_{\mathrm{c}}$ weakens the reconnection trigger $\mathcal{G}$ and reduces the magnetic pressure in the current sheet, $P_{\mathrm{m}}\propto P_{\mathrm{r}}$. Consequently, both the effective event rate and the peak event power are suppressed. This is the gap-choking effect: the migration trap enhances the supply of sBHs, but the same pile-up can deplete the gas required to power strong magnetic reconnection.

This feedback makes the trap a self-regulated heating site rather than a steady heat source. The strongest heating does not necessarily occur exactly at the maximum of the sBH surface density. Instead, it tends to arise near the trap edges or in adjacent regions where the sBH abundance is enhanced but the gas has not been completely depleted. This point is important for interpreting the later light curves: the observable variability is set by the competition between the number of reconnection sites and the amount of gas available to feed them.

Overall, Figure~\ref{fig3} shows that the radial structure of sBH-driven heating is controlled by three coupled effects: the migration-induced sBH pile-up, the gap-induced reduction of the local gas density, and the strength of the underlying viscous disk. This radial heating profile sets the spatial pattern of temperature perturbations that will be evolved in the time-dependent thermal calculation and converted into synthetic optical/UV light curves in the following sections. Because $\Phi\lesssim0.1$ throughout the parameter space explored here, the sBH reconnection heating is energetically subdominant to the viscous background, so treating the background disk structure as fixed during the population evolution is internally self-consistent. We emphasize, however, that this energetic subdominance does not preclude a strong variability imprint. Because the geometrically extended disk acts as a low-pass filter (Figure~\ref{fig4}), the smooth CHAR background carries little power at the shortest time separations, so a spatially localized and temporally bursty heating source can dominate the high-frequency structure function and the inter-band phase lags even while contributing only $\lesssim10\%$ of the local heating budget. The diagnostics presented below are therefore sensitive to the short-timescale, spatially concentrated character of the sBH heating rather than to its absolute energy contribution.

\subsection{Time-domain Variability and Thermal Response}
\label{subsec:time_domain}

\begin{figure*}
    \centering
    \includegraphics[width=0.45\textwidth]{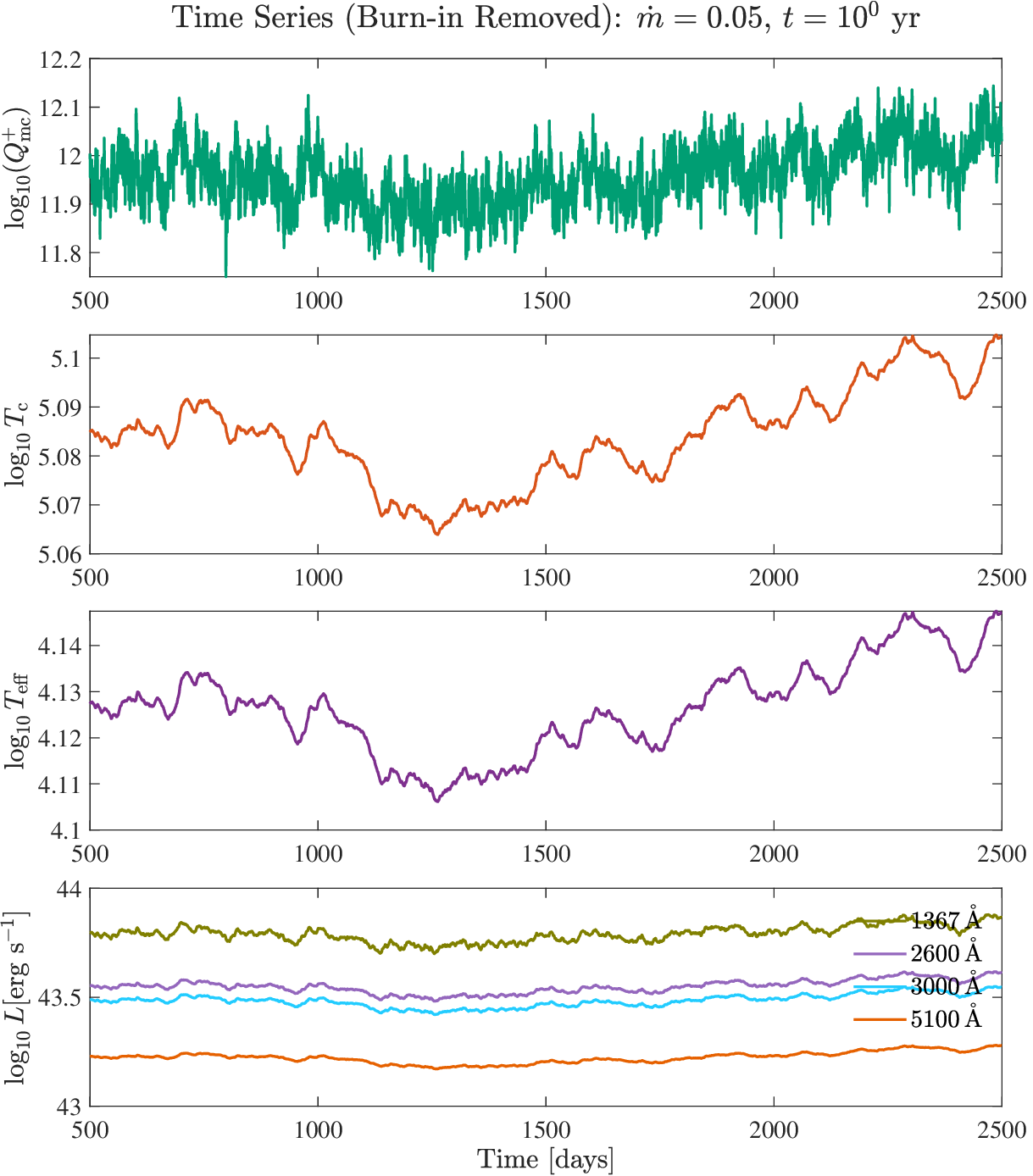}
    \includegraphics[width=0.45\textwidth]{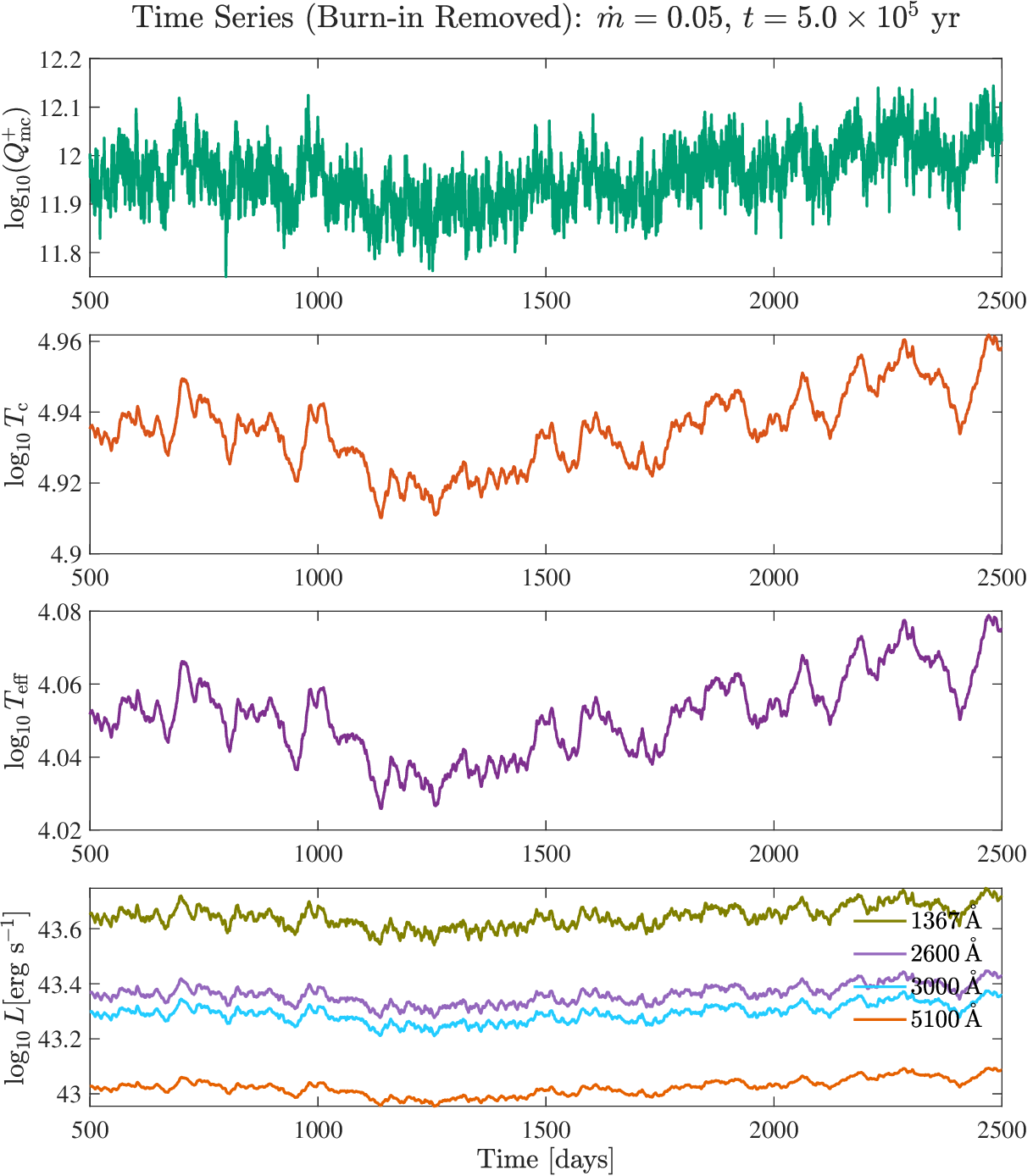}
    \caption{Temporal evolution of the disk thermal response and multi-band light curves at $\dot{m}=0.05$. The left panel shows an sBH-quiescent stage, while the right panel shows a late sBH-active stage after the embedded population has accumulated near the migration trap. From top to bottom, the panels show the background magnetic heating $Q_{\mathrm{mc}}^{+}$, the mid-plane temperature $T_{\mathrm{c}}$, the effective temperature $T_{\mathrm{eff}}$, and the synthetic light curves at $1367$, $2600$, $3000$, and $5100$~\AA{} (for clarity, four of the five computed bands are shown; the $1928$~\AA{} band is omitted). The same background fluctuation is used in both cases. The late-stage model develops sharper temperature perturbations and stronger short-timescale optical/UV variability because localized sBH reconnection heating is added to the disk thermal equation.}
    \label{fig4}
\end{figure*}

Having established the radial structure of the sBH-driven heating, we now examine how this heating is converted into observable optical/UV variability. The structured sBH heating enters the thermal equation through the total heating rate defined in Equation~\eqref{eq37}. The radial profile of $Q_{\mathrm{sBH}}^{+}$ is therefore not spatially uniform: it is enhanced near migration-induced sBH pile-ups, but can be partially suppressed in strongly depleted gap regions. Once inserted into the thermal equation, this structured heating produces localized perturbations in $T_{\mathrm{c}}(R,t)$ and $T_{\mathrm{eff}}(R,t)$, which are then mapped into multi-band optical/UV light curves.

To isolate this effect, Figure~\ref{fig4} compares two representative cases at a fixed accretion rate, $\dot{m}=0.05$. The same background magnetic fluctuation $Q_{\mathrm{mc}}^{+}$ is used in both panels, while the imported sBH population corresponds to different evolutionary stages. The left panel represents the sBH-quiescent limit, in which the sBH contribution is negligible, and the disk variability is dominated by the smooth CHAR-like background. The right panel represents a late sBH-active stage, where the evolved population has accumulated near the migration trap and drives localized magnetic reconnection events.

In the sBH-quiescent phase, the disk behaves like a geometrically extended low-pass filter \cite[e.g.,][]{2020ApJ...891..178S}. The imposed background fluctuation $Q_{\mathrm{mc}}^{+}$ produces only smooth variations in $T_{\mathrm{c}}$ and $T_{\mathrm{eff}}$, and the resulting UV/optical light curves show low-amplitude, slowly varying behavior. This occurs because the blackbody emission at a given wavelength is integrated over a broad range of radii, so incoherent local fluctuations are strongly averaged out.

The late sBH-active phase is qualitatively different. Localized reconnection heating introduces sharp perturbations into the thermal structure, especially near the radii where the sBH heating ratio $\Phi$ is large. These perturbations appear as spikes in both $T_{\mathrm{c}}$ and $T_{\mathrm{eff}}$, and they survive the radial integration into the observed light curves. The resulting UV/optical emission contains short-timescale burst-like features superposed on the smoother CHAR background. Thus, the migration trap affects the light curves indirectly: it first reshapes the radial distribution of $Q_{\mathrm{sBH}}^{+}$, and this structured heating then imprints high-frequency variability on the thermal emission. The same short-timescale excess is also present in the PSD as enhanced high-frequency power, although we focus below on SF diagnostics because they are more directly applicable to irregularly sampled survey data.

\subsubsection{Statistical Properties of Model Light Curves}
\label{subsubsec:state_properties}

\begin{figure*} 
\centering 
\includegraphics[width=0.85\textwidth]{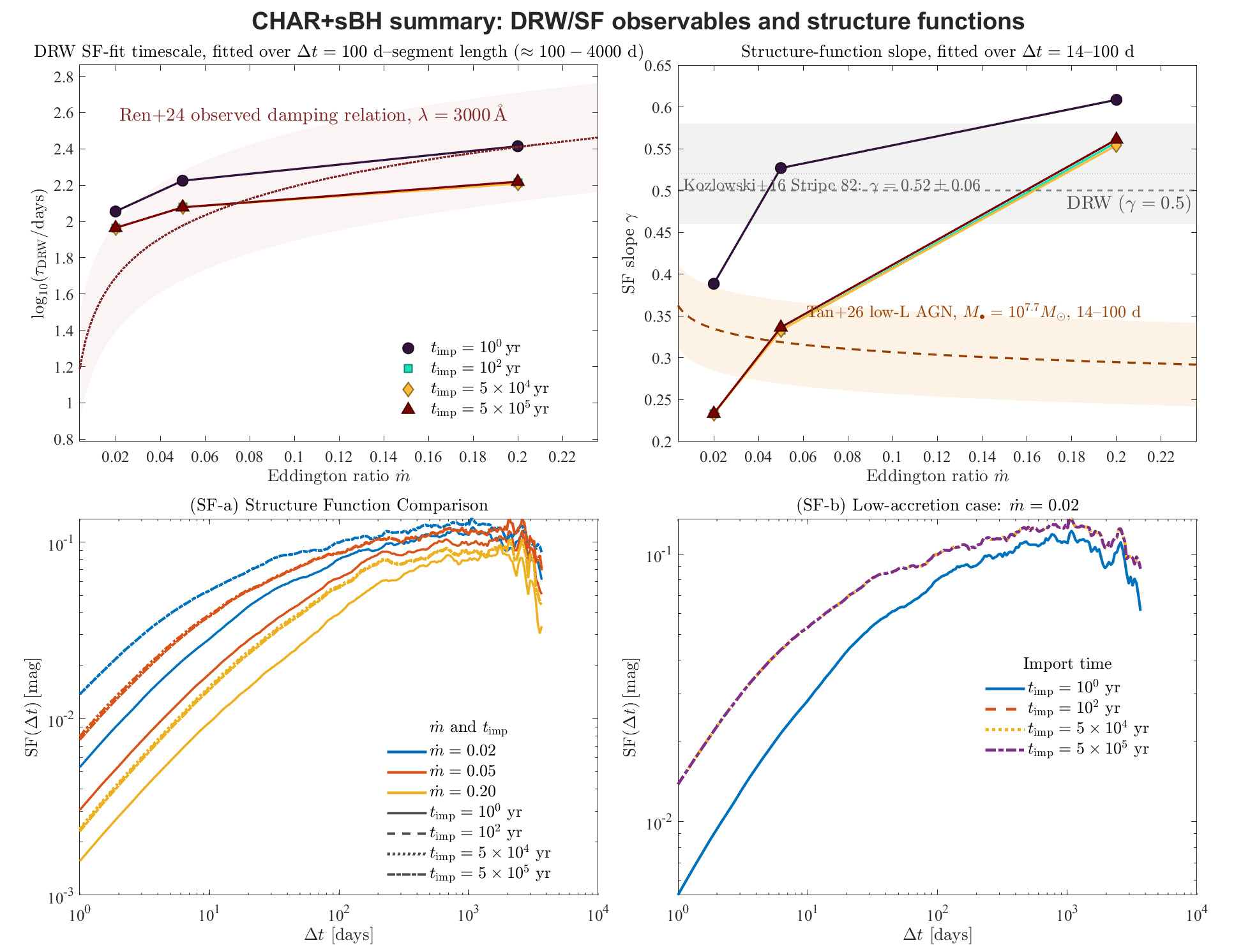} 
\caption{Statistical variability diagnostics of the simulated 3000~\AA{} light curves as functions of Eddington ratio and imported sBH population age. The upper-left panel shows the effective DRW damping timescale obtained by fitting the DRW structure-function form over $\Delta t\approx100$--$4000$ days (100 days up to the per-segment maximum). The red dotted curve and shaded region show the observed \cite{2024ApJ...975..160R} relation and its scatter at $\lambda=3000$~\AA{}. The upper-right panel shows the structure-function slope $\gamma$ measured from a \cite{2020ApJ...891..178S}-style NMAD SF over $\Delta t=14$--$100$~days. The dashed horizontal line marks the short-time DRW expectation $\gamma=0.5$, the gray band shows the Stripe~82 quasar reference range from \cite{2016ApJ...826..118K}, and the orange dashed curve shows the low-luminosity AGN relation from \cite{2026MNRAS.tmp..992T} for comparison. The lower panels show the full SFs, where $\Delta t$ denotes time separation. The $t_{\rm imp}=10^{0}$~yr case is the CHAR-like reference state before sBH-driven reconnection heating becomes important. Evolved sBH populations enhance the short-to-intermediate time-separation variability and flatten the measured SF slope at low and moderate Eddington ratios, while the high-accretion case is more strongly diluted by the brighter background disk.} 
\label{fig5} 
\end{figure*}

We first examine the statistical properties of the synthetic optical/UV light curves before turning to the inter-band delays in Section~\ref{subsubsec:Inter-band time lags}. Figure~\ref{fig5} summarizes the 3000~\AA{} diagnostics as functions of the Eddington ratio $\dot m$ and the imported sBH population age $t_{\rm imp}$. Throughout this subsection $\Delta t$ denotes the time separation entering the SF.

The SF is measured in magnitudes with a segment-averaged normalized median absolute deviation estimator, and the logarithmic SF slope is obtained by fitting
$\log {\rm SF} = \gamma\,\log \Delta t + {\rm const.}$,
over $\Delta t=14$--$100$~days for the 3000~\AA{} light curve. This window is chosen to match the $\Delta t = 14$--$100$ day range over which \cite{2026MNRAS.tmp..992T} fitted their low-luminosity AGN SF slopes, so that our model $\gamma$ can be compared with theirs at the same time separations.

The $t_{\rm imp}=10^{0}$~yr case provides a CHAR-like reference state, since the embedded sBH population has not yet built up significant reconnection heating. In this limit the SF slope is broadly quasar-like, with $\gamma\simeq0.38$, $0.53$, and $0.61$ at $\dot m=0.02$, $0.05$, and $0.20$, respectively. Once the population is imported ($t_{\rm imp}\gtrsim10^{2}$~yr), the low- and moderate-accretion models flatten markedly: $\gamma$ drops to $\simeq0.23$ at $\dot m=0.02$ and to $\simeq0.34$--$0.35$ at $\dot m=0.05$, while the high-accretion case remains steep, $\gamma\simeq0.55$. The flattening arises because localized sBH-triggered reconnection injects additional short-time-separation power, raising the SF at small $\Delta t$ and thereby reducing the log--log slope over $14$--$100$~days. The lower panels of Figure~\ref{fig5} show this directly: at $\dot m=0.02$ the evolved sBH curves lie above the $t_{\rm imp}=10^{0}$ reference at short and intermediate $\Delta t$ and reach the broad plateau earlier, whereas the three evolved ages ($10^{2}$, $5\times10^{4}$, $5\times10^{5}$~yr) are nearly indistinguishable, indicating that once trap-associated heating is established the 3000~\AA{} SF is set by the presence of localized stochastic heating rather than by the precise snapshot age. At $\dot m=0.20$ the brighter, hotter background disk and the larger 3000~\AA{} emitting area dilute the localized events, so the SF slope stays close to the CHAR-like value. The SF slope is thus the most sensitive single diagnostic of localized sBH heating in the low-to-moderate-$\dot m$ regime.

A pure DRW has a short-$\Delta t$ SF slope of $\gamma=0.5$. Even the CHAR-like reference is not a clean single-timescale DRW: while $\dot m=0.05$ is nearly DRW-like ($\gamma\simeq0.53$), it already departs toward shallower ($0.38$ at $\dot m=0.02$) and steeper ($0.61$ at $\dot m=0.20$) effective slopes. The evolved sBH cases push $\gamma$ well below $0.5$ at low and moderate $\dot m$, so those light curves are clearly not describable by a single damping timescale. Two observational references are shown in the upper-right panel of Figure~\ref{fig5} to bracket the relevant range of optical SF slopes. The gray horizontal band is the single-power-law slope $\gamma=0.52\pm0.06$ obtained for $\sim9000$ SDSS Stripe~82 quasars by \citet{2016ApJ...826..118K}; we use it only as an ensemble, population-level reference for the DRW-like slope regime of luminous quasars, not as a source-matched comparison, since those quasars are substantially more massive ($M_{\bullet}\sim10^{8}$--$10^{9}M_\odot$) and more luminous than our fiducial $M_{\bullet}=5\times10^{7}M_\odot$ and are sampled predominantly in the rest-frame UV. The orange dashed curve is the low-luminosity AGN relation of \citet{2026MNRAS.tmp..992T}, $\gamma = 0.11\,\mathcal{M}_{7} - 0.04\log_{10}\dot m + 0.19$ (with $\mathcal{M}_{7}=\log_{10}[M_{\bullet}/10^{7}M_\odot]$), evaluated at $M_{\bullet}=10^{7.7}M_\odot$ with an indicative $\pm0.05$ scatter band. This sample overlaps our parameter range in SMBH mass and Eddington ratio (median $M_\bullet \sim 10^7\,M_\odot$, $\dot m \sim 0.1$). Because we now fit our model slope over the same $14$--$100$ day window, the comparison is matched in time separation; the principal remaining mismatch is in wavelength, since the \cite{2026MNRAS.tmp..992T} slope was measured in the ATLAS-$o$ band, i.e. at a rest-frame wavelength of $\sim$6000--6500\,\AA\ for the sample median redshift rather than at our 3000\,\AA, and carries no explicit wavelength dependence. Overplotting it against our 3000\,\AA\ diagnostic therefore implicitly assumes a wavelength-independent slope, so the comparison remains qualitative. Both reference samples are, moreover, optically selected and more luminous than our fiducial nucleus: the \citet{2016ApJ...826..118K} Stripe~82 quasars span a broad range of Eddington ratios ($\dot m \sim 0.01$--$1$, with a parent-sample median of order $0.1$ \cite[e.g.,][]{2011ApJS..194...45S}) and are used only as a population-level $\gamma$ band rather than at a single $\dot m$. Both samples thus sit near the upper end of our explored Eddington-ratio range; because the evolved-sBH flattening is strongest at the lowest ratio ($\dot m = 0.02$), the most decisive form of this test requires low-$\dot m$ AGN, which are sparsely represented in these luminous samples.

Although the SF is not strictly DRW, it is useful to ask what timescale a forced DRW fit returns, since this is the quantity reported observationally. Fitting the DRW SF from Equation~\eqref{eq45} over $\Delta t = 100$ days up to the per-segment maximum ($\approx4\times10^{3}$ days) yields $\log_{10}(\tau_{\rm DRW}/{\rm day})\simeq1.97$--$2.41$, i.e.\ $\tau_{\rm DRW}\approx90$--$260$~days, for all models in Figure~\ref{fig5}. The CHAR-like reference cases have slightly longer effective timescales than the evolved sBH cases, especially at low and moderate $\dot m$, because the added reconnection heating shifts power to shorter $\Delta t$. The red dotted curve and shaded region in the upper-left panel show the observed damping-timescale relation of \citet{2024ApJ...975..160R}, $\log_{10}(\tau/{\rm day}) = 0.72\log_{10}L_{\rm bol} + 1.19\log_{10}(\lambda/{\rm\AA}) - 34.2$, evaluated at $\lambda=3000$~\AA{} and $M_{\bullet}=5\times10^{7}M_\odot$ with $L_{\rm bol}=\dot m\, L_{\rm Edd}$ and $L_{\rm Edd}=1.26\times10^{38}(M_{\bullet}/M_\odot)~{\rm erg\,s^{-1}}$; the shaded region is a conservative $\pm0.3$~dex guide rather than a formal confidence interval. This relation is well matched to our setup in SMBH mass and luminosity, as it was calibrated on local Seyfert~1 nuclei whose $L_{\rm bol}$ spans our $\dot m\,L_{\rm Edd}$ range, but two points should be kept in mind. First, its wavelength index ($1.19$) is adopted from the CHAR-based prediction of \citet{2024ApJ...966....8Z} because the ZTF $zg/zr$ bands used by \citet{2024ApJ...975..160R} are too close in wavelength to constrain it, so evaluating the relation at $3000$~\AA{} is an extrapolation of the fitted $L_{\rm bol}$ normalization to a bluer band. Second, because this relation was itself shown to be consistent with the CHAR prediction, the agreement here is in part a self-consistency check of the underlying variability framework rather than a fully independent observational test. To assess how strongly this self-consistency limits the comparison, we note that the only CHAR-dependent ingredient is the wavelength index (1.19) used to extrapolate the \citet{2024ApJ...975..160R} normalization to 3000\,\AA. An independent, purely observational constraint on this index is provided by \cite{2025A&A...695A.268S}, who constructed ultraviolet ensemble structure functions from SDSS-RM spectroscopy and derived $\tau \propto L^{0.50\pm0.03}\,\lambda_{\rm RF}^{1.42\pm0.09}$ directly from the SF normalization, without any DRW fit. Because their timescale is read directly from the SF, this measurement is methodologically closer to our own estimator than a DRW fit, and its wavelength index brackets the CHAR-based value of 1.19; adopting it in place of 1.19 shifts the 3000\,\AA\ evaluation of the \citet{2024ApJ...975..160R} relation by only $\lesssim 0.1$\,dex, well within the plotted scatter. The damping-timescale comparison is therefore not critically dependent on the CHAR-derived index. We caution that the \cite{2025A&A...695A.268S} sample comprises luminous quasars ($L_{\rm bol}=10^{45.4}$--$10^{47.1}\,{\rm erg\,s^{-1}}$), more luminous than our fiducial nucleus, and that their $\tau$ carries an arbitrary normalization, so it constrains the wavelength scaling rather than the absolute timescale. With these caveats, the model timescales fall within, or close to, the observed relation, so the key statistical imprint of embedded sBHs is not an anomalous damping timescale---$\tau_{\rm DRW}$ remains observationally reasonable---but rather the combination of enhanced short-$\Delta t$ power and a flatter effective SF slope in the low-to-moderate-$\dot m$ regime. This supports a picture in which migration traps act as localized variability attractors, modifying the optical/UV SF without producing a simple single-timescale DRW.

\subsubsection{Inter-band Time Lags}
\label{subsubsec:Inter-band time lags}

\begin{figure*}
    \centering
    \includegraphics[width=0.85\textwidth]{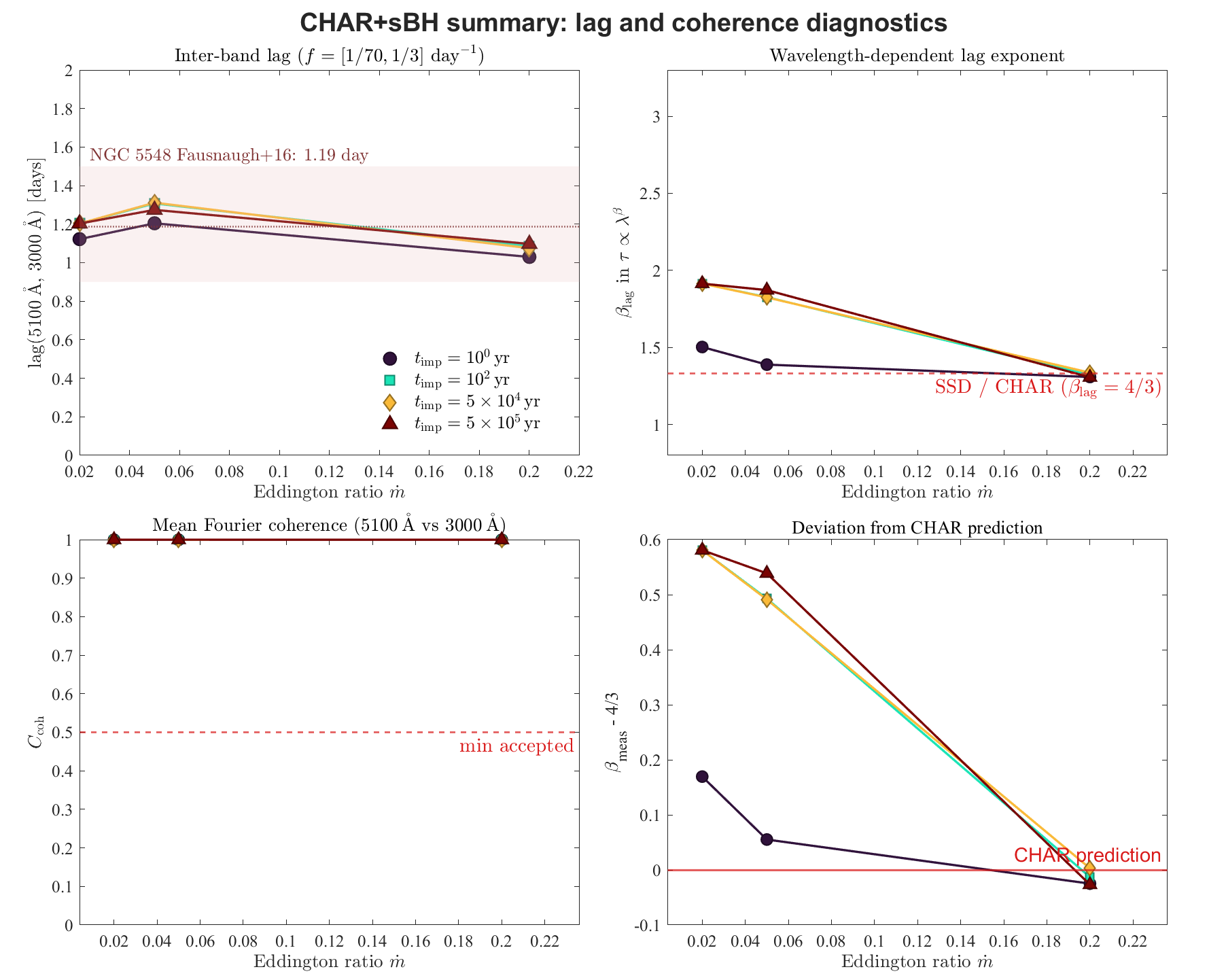}
    \caption{Inter-band lag and coherence diagnostics for the simulated continuum light curves as functions of the Eddington ratio $\dot m$ and the imported sBH population age $t_{\rm imp}$. Marker style encodes $t_{\rm imp}$: $10^{0}$ (circles), $10^{2}$ (squares), $5\times10^{4}$ (diamonds), and $5\times10^{5}$~yr (triangles); the $t_{\rm imp}=10^{0}$~yr case is the CHAR-like reference state. \emph{Upper left}: lag between 5100~\AA{} and 3000~\AA{}, measured from the coherence-weighted phase-lag spectrum over $1/70<f<1/3~\mathrm{day}^{-1}$. The dotted line and shaded band show the NGC~5548 reference, $\tau(5100)-\tau(3000)\simeq1.19$~days, obtained from the no-$U$/$u$ continuum lag-spectrum fit of \citet{2016ApJ...821...56F}, we adopt a conservative band of $0.90$--$1.50$~days to allow for fit uncertainty, $\alpha_{\rm fit}$-$\beta_{\rm fit}$ covariance, and possible diffuse-continuum contamination. \emph{Upper right}: wavelength-dependent lag exponent $\beta_{\rm lag}$ obtained by fitting $\tau(\lambda)\propto[(\lambda/\lambda_0)^{\beta_{\rm lag}}-1]$ with $\lambda_0=1367$~\AA{}; the dashed line marks the standard thin-disk / CHAR value $\beta_{\rm lag}=4/3$. \emph{Lower left}: mean Fourier coherence $C_{\rm coh}$ between the two bands; the dashed line is the minimum accepted value, $C_{\rm coh}=0.5$. \emph{Lower right}: deviation $\beta_{\rm lag}-4/3$ from the thin-disk prediction. Evolved sBH populations steepen $\beta_{\rm lag}$ above $4/3$ at low and moderate $\dot m$ while keeping the absolute $5100$/$3000$~\AA{} lag within the observed band, and the effect washes out at $\dot m=0.20$.}
    \label{fig6}
\end{figure*}

We now turn to the continuum inter-band delays, measured with the frequency-domain phase-lag method (Figure~\ref{fig6}), where a positive lag means that the longer-wavelength band lags the shorter-wavelength reference band.

The $5100$~\AA{} versus $3000$~\AA{} delay, averaged over $1/70<f<1/3~\mathrm{day}^{-1}$, is $\simeq1.0$--$1.3$~days across all models, with the CHAR-like reference at $\simeq1.0$--$1.2$~days and the evolved cases marginally longer at low and moderate $\dot m$. For comparison, the dotted line and shaded band in the upper-left panel give the NGC~5548 reference: adopting the no-$U$/$u$ continuum lag-spectrum fit of \citet{2016ApJ...821...56F}, $\tau(\lambda)=\alpha_{\rm fit}\,[(\lambda/\lambda_0)^{\beta_{\rm fit}}-1]$ with $\alpha_{\rm fit}=0.79$~day, $\beta_{\rm fit}=0.99$, and $\lambda_0=1367$~\AA{}, gives $\tau(5100)-\tau(3000)\simeq1.19$~days; we plot a conservative band of $0.90$--$1.50$~days to allow for fit uncertainty, $\alpha_{\rm fit}$-$\beta_{\rm fit}$ covariance, and possible diffuse-continuum contamination (the \citet{2016ApJ...821...56F} measurement is consistent with, but of higher quality than, that of \citealt{2015ApJ...806..129E}). Unlike the SF-slope reference bands above, this is a genuinely source-matched comparison: the reverberation-mapped mass of NGC~5548, $M_{\bullet}=(5.2\pm1.3)\times10^{7}M_\odot$, and its accretion rate, $L_{\rm bol}\approx0.1\,L_{\rm Edd}$, are close to our fiducial $M_{\bullet}=5\times10^{7}M_\odot$ and to the $\dot m\approx0.05$--$0.2$ models. The comparison is nonetheless made at the level of the characteristic lag amplitude rather than with an identical estimator, since our value is a frequency-domain phase lag averaged over timescales of a few to $\sim70$~days, whereas the \citet{2016ApJ...821...56F} value derives from a time-domain cross-correlation/JAVELIN analysis of the full campaign. With this in mind, all model lags fall within the observed band, and in particular the CHAR-like reference reproduces the observed delay, consistent with the ability of the CHAR framework to recover AGN continuum lags \citep{2020ApJ...891..178S}; the absolute $5100$/$3000$~\AA{} lag therefore does not, by itself, discriminate strongly between the reference and evolved states.

The discriminating signature is the wavelength scaling. Fitting $\tau(\lambda)\propto[(\lambda/\lambda_0)^{\beta_{\rm lag}}-1]$ with $\lambda_0=1367$~\AA{}, the CHAR-like reference gives $\beta_{\rm lag}\simeq1.4$--$1.5$, close to the standard thin-disk and CHAR expectation $\beta_{\rm lag}=4/3$ (dashed line) \cite[e.g.,][]{2014MNRAS.444.1469M,2015ApJ...806..129E,2021iSci...24j2557C,2021ApJ...907...20K}. We further note that the observed continuum lag--wavelength slopes lie at or below $4/3$ (e.g.\ $\beta_{\rm fit}\simeq0.99$ for the free-$\beta$ fit of \citealt{2016ApJ...821...56F}, statistically consistent with $4/3$ within the uncertainties), so the evolved-population value $\beta_{\rm lag}\simeq1.9$ at low $\dot m$ represents a clear, testable departure above the observed range, rather than merely a shift within it. The evolved sBH cases steepen the relation to $\beta_{\rm lag}\simeq1.9$ at $\dot m=0.02$ and $\simeq1.8$--$1.9$ at $\dot m=0.05$, before returning to $\simeq1.33$ at $\dot m=0.20$; equivalently, the deviation $\beta_{\rm lag}-4/3$ (lower-right panel) peaks at $\simeq0.55$--$0.58$ at low $\dot m$ and collapses to $\approx0$ at high $\dot m$. The steepening occurs because the radially localized sBH heating does not follow the smooth viscous temperature profile and distorts the disk thermal response. We caution that the precise value of $\beta_{\rm lag}$ is calibration- and $t_{\rm imp}$-dependent and is not a source-by-source prediction; the robust, falsifiable feature is the sign and accretion-rate dependence: localized sBH heating drives $\beta_{\rm lag}>4/3$, the effect is strongest at low-to-moderate $\dot m$, and it weakens toward the thin-disk value as $\dot m$ increases.

The mean Fourier coherence $C_{\rm coh}$ stays close to unity and well above the adopted minimum of $0.5$ (lower-left panel) for all reliable measurements, so the recovered delays are not dominated by incoherent shot noise, and the $\beta_{\rm lag}$ steepening at low and moderate $\dot m$ reflects a genuine change in the disk thermal response. At $\dot m=0.20$ the brighter background and larger emitting area average over the localized perturbations, returning $\beta_{\rm lag}$ to the thin-disk value; the absence of a lag anomaly at high $\dot m$ therefore does not imply the absence of embedded sBHs, but rather their dilution by the smooth disk emission. Overall, the most favorable regime for detecting embedded-sBH signatures is the low-to-moderate Eddington-ratio range, where the continuum lag amplitude remains observationally consistent while the lag--wavelength relation steepens measurably above the standard thin-disk prediction.

\section{Discussion and Conclusions}
\label{sec:discussion and conclusion}

\subsection{Parameter Dependence and Model Limitations}
\begin{figure}
    \centering
    \includegraphics[width=0.48\textwidth]{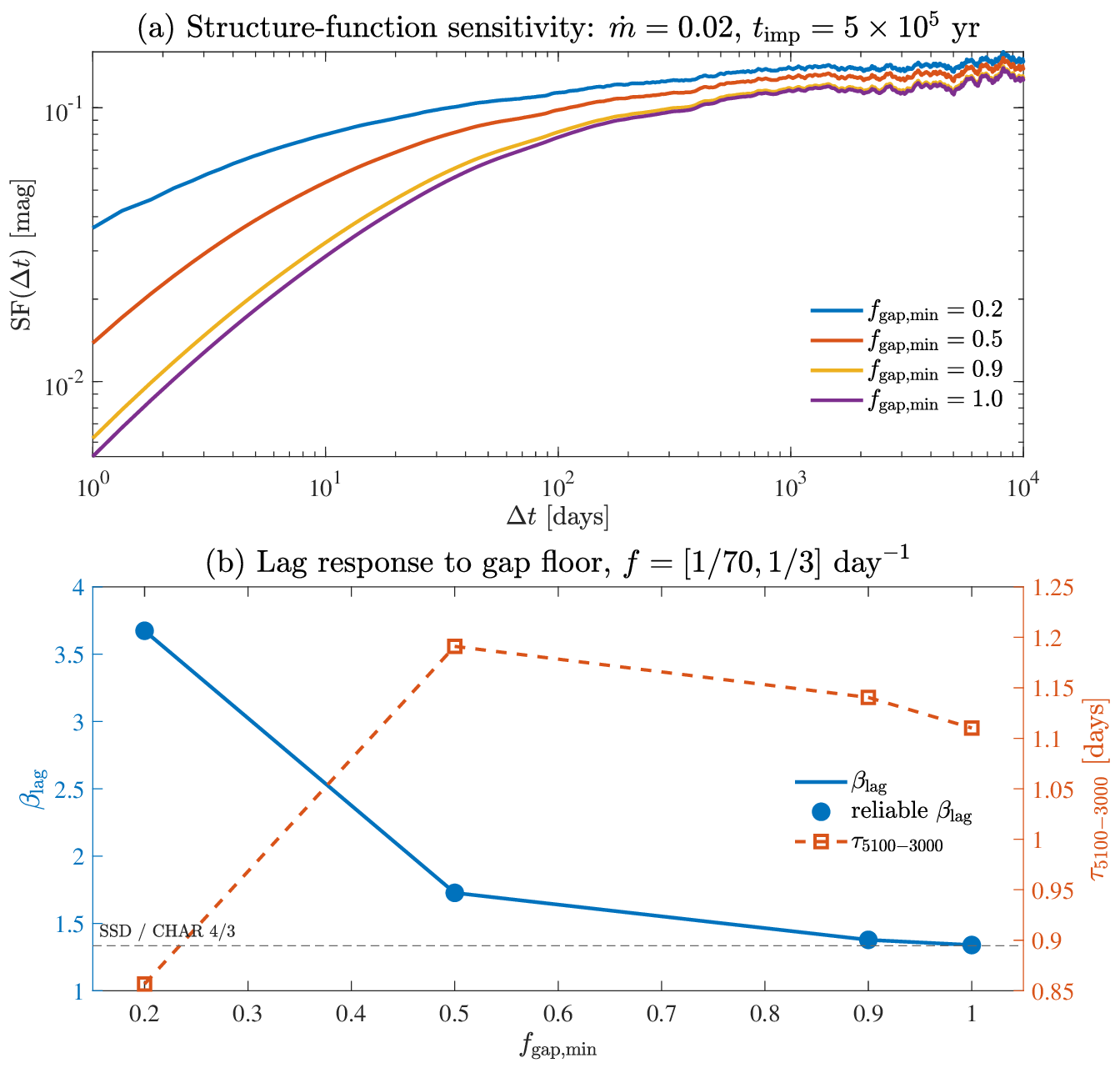}
    \caption{Sensitivity of the variability diagnostics to the gap-depletion floor $f_{\rm gap,min}$, for a representative low-accretion, late-time model ($\dot m=0.02$, $t_{\rm imp}=5\times10^{5}$~yr). (a) 3000~\AA{} structure function for $f_{\rm gap,min}=0.2$, $0.5$, $0.9$, and $1.0$. A deeper gap (smaller $f_{\rm gap,min}$) lowers the thermal inertia of the depleted annulus and raises the SF at short and intermediate time separations, while all models converge to a similar asymptotic amplitude at $\Delta t\gtrsim10^{3}$~days. (b) Corresponding lag response measured over $1/70<f<1/3~\mathrm{day}^{-1}$: the wavelength-dependent exponent $\beta_{\rm lag}$ (blue, left axis) and the absolute $5100$/$3000$~\AA{} lag $\tau_{5100-3000}$ (orange, right axis) as functions of $f_{\rm gap,min}$. The dashed line marks $\beta_{\rm lag}=4/3$. The exponent $\beta_{\rm lag}$ decreases monotonically with $f_{\rm gap,min}$, whereas $\tau_{5100-3000}$ is non-monotonic, peaking near the observed NGC~5548 value at an intermediate gap floor.}
    \label{fig7}
\end{figure}

The results presented above should be interpreted in light of several model parameters that control the strength, intermittency, and radial localization of the sBH-induced heating. Among them, the gap-depletion factor is particularly important. In our treatment, the one-dimensional gap factor $f_{\rm gap}^{\rm 1D}$, together with the imposed floor $f_{\rm gap,min}$, regulates both the local gas supply available for reconnection and the thermal inertia of the depleted disk annulus. A smaller $f_{\rm gap,min}$ corresponds to a deeper effective gap. This reduces the co-moving gas density and therefore tends to weaken the ram-pressure-driven reconnection trigger, but it also lowers the local surface density, optical depth, and vertically integrated heat capacity, allowing a given heating perturbation to produce a larger fractional temperature response. Conversely, a larger $f_{\rm gap,min}$ makes the gap shallower, maintains a larger gas reservoir and a smoother reconnection rate, but also increases the thermal inertia of the disk. The observable response is therefore not expected to vary monotonically with $f_{\rm gap,min}$; it depends on the competition between gap choking of the reconnection power and gap-assisted amplification of the local thermal response.

Figure~\ref{fig7} makes this competition explicit for a representative low-accretion, late-time model ($\dot m=0.02$, $t_{\rm imp}=5\times10^{5}$~yr). Panel~(a) shows that a deeper gap raises the 3000~\AA{} structure function at short and intermediate time separations: as $f_{\rm gap,min}$ decreases from $1.0$ to $0.2$, ${\rm SF}(\Delta t)$ at $\Delta t\sim1$--$10$~days increases by nearly an order of magnitude, because the reduced thermal inertia of the depleted annulus amplifies the fractional temperature response to each reconnection burst. All curves nonetheless converge to a similar asymptotic amplitude, ${\rm SF}_{\infty}\sim0.15$~mag, at $\Delta t\gtrsim10^{3}$~days, so the gap floor primarily controls the short-time-separation SF shape rather than the long-term variance. Panel~(b) shows the corresponding lag response. The wavelength-dependent exponent decreases monotonically with the gap floor, from $\beta_{\rm lag}\simeq3.7$ at $f_{\rm gap,min}=0.2$ to the thin-disk value $\beta_{\rm lag}\simeq4/3$ at $f_{\rm gap,min}=1.0$, confirming that deeper gaps produce steeper, more anomalous lag--wavelength relations. The absolute $5100$/$3000$~\AA{} lag, by contrast, is non-monotonic: it is shortest ($\simeq0.86$~days, just below the observed NGC~5548 band) for the deepest gap, peaks near the observed value ($\simeq1.2$~days) at $f_{\rm gap,min}=0.5$, and declines only mildly toward shallow gaps. This is precisely the competition anticipated above: the deepest gap maximizes the local thermal amplification—and hence the steepness of $\beta_{\rm lag}$—but simultaneously chokes the reconnection power and shortens the effective inter-band delay, whereas an intermediate gap floor ($f_{\rm gap,min}\simeq0.5$, close to the fiducial value adopted in Figures~\ref{fig5} and~\ref{fig6}, where $\tau_{5100-3000}\simeq1.2$~days) simultaneously yields a steepened $\beta_{\rm lag}$ and an absolute lag consistent with NGC~5548. In the fiducial set of simulations explored here, the strongest signatures appear in low-to-moderate Eddington-ratio models after the embedded population has had time to accumulate near migration traps. This should not be interpreted as a full survey of the multidimensional parameter space. The observable amplitude remains degenerate with the sBH supply rate, mass function, gap floor, reconnection efficiency, and thermal-coupling parameters.

Other parameters affect the normalization and stochasticity of the signal in partly degenerate ways. The effective pressure ratio $\beta_\mathrm{p}=P_{\mathrm{r}}/P_{\mathrm{m}}$, the magnetization $\sigma$, the current-sheet aspect ratio $g_s$, and the interior deposition fraction $f_{\rm int}$ set the amount of reconnection power that is ultimately coupled to the disk interior. The lognormal width $\sigma_{\rm amp}$ controls the burst-to-burst intermittency and therefore mainly affects short-lag SF amplitudes and the apparent DRW damping time. Population parameters such as $\dot N_{\rm cap}$, $\eta_{\rm BH}$, and the sBH mass-function slope determine the number and mass of embedded objects that reach the migration trap. Since $K'\propto q^2 h^{-5}\alpha^{-1}$, the trap structure is especially sensitive to the characteristic sBH mass, disk aspect ratio, and viscosity parameter. Varying these parameters can shift the trap radius, change the degree of pile-up, and alter the relative importance of the pile-up and gap-opening effects. A full parameter exploration is therefore required before the predicted amplitudes can be mapped directly onto observed AGN samples. In particular, throughout this work we fix the SMBH mass at $M_\bullet=5\times10^{7}M_\odot$ and vary only the Eddington ratio and the imported population age; a systematic study of the mass dependence, which sets both the disk temperature profile and the absolute lag normalization, is left to future work. The present calculations should thus be read as a controlled, single-mass proof of concept rather than as a population-level prediction.

A further caveat is that the present calculation is not fully self-consistent in the global disk-evolution sense. The embedded sBH population is evolved on a prescribed background disk, and the resulting sBH distribution is then coupled to the CHAR-like thermal equation as an additional localized heating source. We include the local effect of gap depletion on the co-moving gas density and on the thermal quantities entering the disk response, but we do not allow the sBH heating, accretion, or gap opening to feed back onto the global surface-density profile, mass accretion rate, disk aspect ratio, opacity structure, or migration torques. Such feedback could be important. In AGN disks, radiative or mechanical feedback from embedded compact objects can modify the local gas supply, suppress accretion, excavate underdense cavities, and alter the disk temperature structure and mass flux \citep[e.g.,][]{2022ApJ...927...41T,2023ApJ...948..136C,2025MNRAS.537.3396E,2024ApJ...966L...9Z}. Moreover, gap-opening perturbers exert torques that depend on the gas surface density inside the gap, while local heat release can generate thermal torques and change the direction or efficiency of migration \citep[e.g.,][]{2015Natur.520...63B,2017MNRAS.472.4204M,2018ApJ...861..140K}. In a globally coupled disk, local sBH heating may therefore puff up the disk, change $h$, shift $K'\propto q^2 h^{-5}\alpha^{-1}$, and move or weaken the migration trap itself. Similarly, a sufficiently dense embedded population could modify the radial gas flow and the local magnetic environment, thereby changing both the supply of sBHs and the efficiency of reconnection. The present model should therefore be viewed as a controlled proof-of-concept calculation that isolates the observable consequences of sBH pile-ups, rather than as a fully coupled radiation-MHD population-disk simulation.

Future work should relax these assumptions by evolving the gas disk, the embedded compact-object population, and the magnetic heating source in a single coupled framework. In such models, the migration torques and gap depths should be recomputed from the instantaneous disk structure, and the response of the corona and X-ray emission should be included in order to compare simultaneously with optical/UV and X-ray reverberation data. On the observational side, high-cadence multi-band optical surveys will be essential for testing the predicted combination of enhanced short-timescale variability, flattened short-lag SFs, and anomalous lag-wavelength slopes, using only reliable high-coherence lag measurements. The Wide Field Survey Telescope (WFST), with its dedicated optical time-domain survey capability \citep[e.g.,][]{2023SCPMA..6609512W}, provides a particularly promising platform for such tests. Mock WFST light curves generated from the present model, including realistic cadence, depth, seasonal gaps, and photometric noise, would allow us to quantify whether sBH-induced disk heating can be statistically distinguished from ordinary CHAR-like variability and other inhomogeneous-disk fluctuations.

\subsection{Summary and Outlook}
\label{subsec:summary}

In this work, we investigated whether embedded sBHs in AGN disks can leave observable optical/UV variability signatures. The essential physical sequence is that migration traps first concentrate the embedded sBH population; the enhanced local population increases the number of reconnection sites, while gap opening simultaneously reduces the gas density that powers reconnection. The competition between pile-up-enhanced event rates, gap-choked reconnection power, and reduced thermal inertia in depleted annuli therefore controls the observable variability. 

Motivated by theoretical expectations that AGN disks can host compact objects supplied by nuclear-star-cluster capture and in-situ formation \cite[e.g.,][]{2016ApJ...819L..17B,2019ApJ...878...85S,2022ApJ...928..191G,2026ApJ...999...55C}, we combined a standard thin-disk model with a CHAR-like magnetic variability prescription \cite[e.g.,][]{1973A&A....24..337S,2020ApJ...891..178S} and a one-dimensional population synthesis of migrating and accreting sBHs. The evolved sBH distribution was then coupled to a magnetic reconnection heating model, following the idea that compact objects moving through a magnetized AGN disk can trigger localized disk-side energy release \cite[e.g.,][]{2025ApJ...991..167X}. 

Our primary physical result is that migration traps act as attractors for variability. Accretion-modified migration can concentrate sBHs at preferred radii, especially when gap opening and accretion-driven torques alter the standard inward-migration picture \cite[e.g.,][]{2018ApJ...861..140K,2026ApJ...997..160I,2026ApJ...997..161P}. These pile-ups increase the number of potential reconnection sites and enhance localized magnetic heating. However, the same accumulated sBHs can also open gaps and reduce the local gas density, weakening the ram-pressure trigger for reconnection. The resulting heating is therefore self-regulated rather than monotonic: it is controlled by the competition between sBH pile-up, gap choking, and dilution by the smooth background disk emission. When this structured heating is inserted into the disk thermal equation, it produces localized perturbations in $T_{\mathrm{c}}$ and $T_{\mathrm{eff}}$, which appear as burst-like optical/UV variability superposed on the smoother CHAR-like background. At low-to-moderate Eddington ratios, the sBH contribution can enhance short-timescale variability, flatten the short-lag SF, and modify the lag-wavelength relation away from the standard $\tau\propto\lambda^{4/3}$ expectation. These effects are qualitatively relevant to observed AGN variability anomalies, including distorted lag-wavelength relations, enhanced short-timescale variability, and departures from simple DRW-like behavior \cite[e.g.,][]{2014MNRAS.444.1469M,2015ApJ...806..129E,2021Sci...373..789B,2021iSci...24j2557C,2021ApJ...907...20K,2024MNRAS.529.2877M}. 

We stress that the predicted steepening ($\beta_{\rm lag}>4/3$) is a new, distinct signature rather than an explanation of the commonly observed lags, which tend to lie at or below the thin-disk slope; its opposite sign makes it a clean, falsifiable discriminator. At high accretion rates, by contrast, the stronger thermal background and larger emitting area dilute the fractional sBH imprint, causing the observable diagnostics to approach the smooth-disk baseline. 

These results suggest that embedded compact-object populations may be constrained through a joint set of optical/UV diagnostics rather than through any single observable. A combination of enhanced short-timescale variability, flattened short-lag SFs, anomalous lag-wavelength scalings, and a weakening of these signatures in high-$\dot{m}$ systems would be a promising candidate signature of localized sBH-driven disk heating. Future work should extend the present disk-side calculation to include the delayed coronal response and X-ray emission, enabling a unified comparison with multi-wavelength reverberation campaigns. In addition to electromagnetic variability, gravitational wave observations can also serve as an important supplementary probe for the gas-rich environment of compact celestial bodies. Although binary star formation, mergers, and spin evolution are beyond the scope of this model, the recent GWTC-4 gravitational-wave population analysis has identified a high-spin, preferentially aligned hierarchical-merger subpopulation, whose properties may be naturally related to the gas-assisted merger process in AGN disks \cite[e.g.,][]{2026PhRvL.137b1407L}. Therefore, the combined study of electromagnetic and gravitational waves is expected to provide complementary constraints for the compact celestial body populations embedded within AGN disks.

\section*{Acknowledgments}
We thank Bao-Quan Huang and Xiao-Yan Li for helpful discussions. This work was supported by the National Key R\&D Program of China under grant 2023YFA1607902, the National Natural Science Foundation of China under grants 12494572, 12221003, 125B2058, 12373070, 12192223, 12494575, and 12273005.

\clearpage
\end{document}